\documentclass[a4paper,fleqn]{cas-dc}
\usepackage[square, numbers, comma, sort&compress]{natbib}
\usepackage{setspace}
\usepackage{listings}
\usepackage{xcolor}
\usepackage[ruled]{algorithm2e}
\usepackage{graphicx}
\usepackage{amsmath}
\usepackage{amssymb}
\usepackage{mathrsfs}
\usepackage{amsthm}
\usepackage{subfig}
\usepackage{wrapfig, blindtext}
\usepackage[utf8x]{inputenc}
\usepackage{bm}
\usepackage{float}
\usepackage{nccmath}
\usepackage{siunitx}
\definecolor{rev1}{rgb}{1,0,0}
\usepackage{adjustbox}

\def\tsc#1{\csdef{#1}{\textsc{\lowercase{#1}}\xspace}}
\tsc{WGM}
\tsc{QE}
\tsc{EP}
\tsc{PMS}
\tsc{BEC}
\tsc{DE}

\begin{document}
\let\WriteBookmarks\relax
\def\floatpagepagefraction{1}
\def\textpagefraction{.001}
\shorttitle{DRL based controller for AUV}
\shortauthors{ST Havenstrom et~al.}
\title [mode = title]{Proportional integral derivative controller assisted reinforcement learning for path following by autonomous underwater vehicles}                      
\author[1]{Simen Theie Havenstrom}
\ead{simentheie@gmail.com}
\author[2]{Camilla Sterud}
\ead{camilla.sterud@sintef.no}
\author[1,2]{Adil Rasheed}
\cormark[1]
\ead{adil.rasheed@ntnu.no}
\ead[url]{www.adilrasheed.com}
\address[1]{Department of Engineering Cybernetics, Norwegian University of Science and Technology, Trondheim, Norway}
\address[2]{Mathematics and Cybernetics, SINTEF Digital, Oslo, Norway}
\author[3]{Omer San}
\ead{osan@okstate.edu}
\ead[url]{www.cfdlab.org}
\address[3]{Mechanical and Aerospace Engineering Department, Oklahoma State University, Oklahoma, USA}
\cortext[cor1]{Corresponding author}

\begin{abstract}
    Control theory provides engineers with a multitude of tools to design controllers that manipulate the closed-loop behavior and stability of dynamical systems. These methods rely heavily on insights about the mathematical model governing the physical system. However, if a system is highly complex, it might be infeasible to produce a reliable mathematical model of the system. Without a model most of the theoretical tools to develop control laws break down. In these settings, machine learning controllers become attractive: Controllers that can learn and adapt to complex systems, developing control laws where engineers cannot. This article focuses on utilizing machine learning controllers in practical applications, specifically using deep reinforcement learning in motion control systems for an autonomous underwater vehicle with six degrees-of-freedom. Two methods are considered: end-to-end learning, where the vehicle is left entirely alone to explore the solution space in its search for an optimal policy, and proportional integral derivative controller assisted learning, where the deep reinforcement learning controller is split into three separate parts, each controlling its own actuator.
\end{abstract}

\maketitle

\section{Introduction}
\label{section:introduction}
    Autonomous Underwater Vehicles (AUV) are employed in various subsea applications, such as seafloor mapping, pipeline inspection and research operations \cite{xiang2010coordinated,bogue2015underwater,ventura2018mapping}. The diversity of operational settings for AUVs implies that truly autonomous vehicles must be able to follow spatial trajectories, maintain a desired velocity and avoid collisions. Achieving all of these simultaneously is a daunting control objective. Within the realm of control theory, mathematical proofs of stability and robustness have always been of great importance. Their attractiveness is obvious, as they can provide guarantees for stability margins, disturbance rejection and convergence rates. However, such proofs often rely heavily on mathematical modeling and preexisting knowledge of the physical world. Following this, most of the approaches in traditional control theory demand making simplifications and assumptions about the system at hand. As the underlying physics grow more complex, these assumptions and simplifications compromise the guarantees, and yield inferior efficiency and performance. An alternative to the traditional, purely theoretical approach is to accept the lack of knowledge about the physical world, and embrace an experimental approach. The argument is that if the system is such that one cannot make these guarantees anyway, why not try to expand the knowledge about the system's behaviour though experimental exploration? Here, experimental exploration refers to the process of extracting information from a system by strategically exciting its inputs. The goal of this process is to use this newly gained information for system identification. A well known version of this process is transfer function identification by studying the step-response of a system in the frequency domain. Conducting such experiments can, however, be challenging, expensive and downright dangerous, as one risks driving the system into unsafe areas of operation. In addition, the desired excitation of the system might be infeasible, as inputs can be subjected to physical constraints, and states of interest may not be observable. Hence, real life experiments quickly become infeasible in many applications.
    
    The picture painted here of the polar opposite approaches is obviously an exaggerated one, and undoubtedly most control approaches fall somewhere in the middle of the continuum bridging the two sides. The late surge in the field of machine learning (ML), especially deep learning (DL) can offer an alternate middle ground for control engineers. In recent years, DL has led to dramatic improvements in the state-of-the-art methods in several domains, such as speech recognition, visual object recognition, object detection, and drug discovery \cite{lecun2015deep}. DL has also accelerated the field of reinforcement learning (RL), leading to the combination of the two fields, known as deep reinforcement learning (DRL). In DRL, deep neural networks (DNN) are used to estimate value functions and effectively act as decision makers, called agents \cite{arulkumaran2017deep}. There are many success stories of DRL, where playing Go \cite{silver2016mastering}, dynamical maneuvering of legged robots \cite{hwangbo2019learning}, autonomous helicopter flight \cite{abbeel2007application} and chip placement \cite{mirhoseini2020chip} are representative examples. DRL has also been proposed as an important contribution in the area of optimal control design \cite{lewis2012reinforcement,bucsoniu2018reinforcement}. Further, RL has been used for designing autopilots and navigation systems of autonomous unmanned aerial vehicles (AUAVs) \cite{bou2010controller, pham2018autonomous}. This new approach is part of the field known as machine learning control (MLC) \cite{kostov1995machine,duriez2017machine,quade2020machine}, where the intention is to set up a robust ML scheme that inputs a set of observations and outputs control actions. In this way, a control law can be effectively learned, instead of explicitly designed. Consequently, the approach is a data driven one, but the data could be synthetically generated (for instance by a simulated model), or be collected from physical experiments. 
    
    One of the important control objectives in the current context is path following. The path-following problem is well studied in traditional control literature \cite{fossen2000nonlinear,kim2015integral,xiang2018survey}. For path following, as in the previously mentioned fields, DRL approaches have proven to be robust and successful potential alternatives to traditional algorithms (\cite{zhou2019learn,zhang2019decision,guo2020autonomous}), particularly in complex environments \cite{chen2019knowledge,biferale2019zermelo,yang2020micro}. In \cite{Yu} the authors propose the deep deterministic policy gradients (DDPG) method for solving the trajectory tracking problem, which is demonstrated to out-compete a proportional integral derivative (PID) controller. A target tracking controller using policy search methods was designed in \cite{El-Fakdi}, while in \cite{Cui} the authors improved trajectory tracking performance using two neural networks in an actor-critic configuration to estimate compensation for unknown nonlinearities and disturbances. In our previous works (\cite{PPO-article,meyer2020taa}), we investigated how DRL could solve a two-dimensional path-following and collision avoidance (with stationary obstacles) problem for a simulated autonomous surface vehicle (ASV). Extending the work in \cite{meyer2020colreg}, it was demonstrated that the DRL could also learn path following and collision avoidance with moving obstacles while also respecting collision avoidance regulations (COLREG). These works also demonstrated the robustness of Proximal Policy Optimization (PPO) for these kinds of applications. One downside of these works was that environmental disturbances were not taken into account. However, literature exists which shows that this aspect can also be taken into consideration with great success. For example, a fast marching method has been designed for operation in dynamic environments using the constraints \cite{liu2015path,song2017multi}. Others have proposed to deal with the disturbances by creating extended state observers which are used in disturbance rejecting control laws  \cite{peng2018ofp, peng2019pfc}.
    
    Despite the impressive success of the previous works there are two common shortcomings. Firstly, they all consider motion in the horizontal plane ie. in 3-degrees of freedom (DOF) only and secondly, most of them focus on end-to-end learning which are computationally demanding and not always very stable. To this end, the current work addresses the problem of path following in three dimensions (\cite{encarnacao20003d,lapierre2007nonlinear,yu2017nonlinear}) with full 6DOF for an AUV under the influence of environmental disturbances in the form of ocean currents. Furthermore, we employ a hybrid approach to training where PID assisted training is used to split the control architecture into three separate parts (one for each actuator of the AUV), leading to computationally efficient and stable training of the DRL agent.
    
    The article is organized as follows: In Section \ref{section:theory}, background theory on AUV modeling, the path-following control problem and DRL is given. The implementation of the simulation model and the utilized DRL algorithm are briefly described in Section \ref{section:implementation}. The main innovative contribution is detailed in Section \ref{section:experiments}. We conclude in Section \ref{section:conclusion} through simulations that achieving a well performing 3D path-following controller through DRL is feasible, at least in theory. Finally we end the paper by proposing future research directions to extend the research work presented here.

\section{Theory}
\label{section:theory}
    
    \subsection{Governing equations of motion for an AUV}
    \label{subsec:mda}
        In this section we briefly present the system of equations governing the motion of an AUV. In marine systems modeling, this representation involves a transformation between different reference frames. The notation used in this article to detail the equations of motion for the marine vessel, is called SNAME (1950), elaborated in Table \ref{tab:notation} \cite{Fossen}.  
        \begin{table*}[pos=h]
        \caption{Notation for marine vessels as given by SNAME (1950)}
        \begin{tabular}{||l|c|c|c|c|| }
        \hline
        \textbf{Degree of freedom} & \textbf{Forces and moments} & \textbf{Velocities} & \textbf{Positions} \\
        \hline
        1 translation in the $x_b$ direction (surge)  & $X$ & $u$ & $x$ \\
        2 translation in the $y_b$ direction (sway)   & $Y$ & $v$ & $y$ \\
        3 translation in the $z_b$ direction (heave)  & $Z$ & $w$ & $z$ \\
        4 rotation about $x_b$ axis (roll)        & $K$ & $q$ & $\phi$ \\
        5 rotation about $y_b$ axis (pitch)       & $M$ & $p$ & $\theta$ \\
        6 rotation about $z_b$ axis (yaw)         & $N$ & $r$ & $\psi$ \\
        \hline
        \end{tabular}
        \label{tab:notation}
        \end{table*}
        
\begin{figure}[pos=h]
            \centering
            \includegraphics[width=\linewidth]{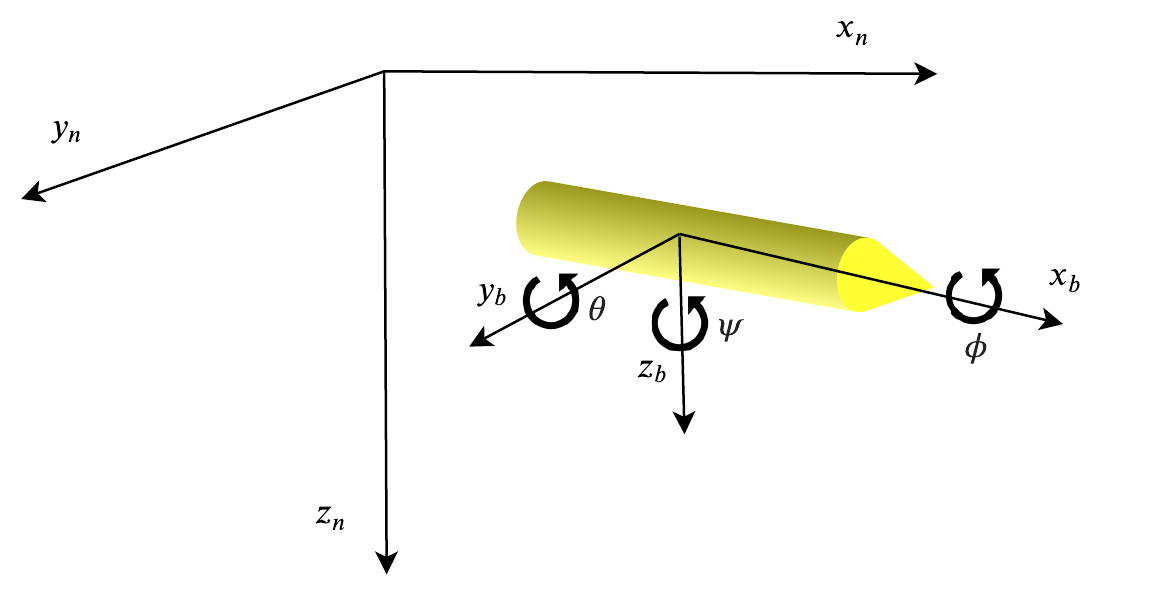}    \caption{Simple illustration of BODY and NED coordinate systems. The BODY frame is obtained by translating and rotating the NED frame by its principal axes.}
            \label{fig:ref_frames}
        \end{figure}

        Modeling motion dynamics for an AUV involves transformation between coordinate systems. The two coordinate systems of interest are the body-frame, $\{b\}$, which is the body-fixed reference frame with origin at the vessel's center of control (CO), and the North-East-Down (NED) coordinate system, $\{n\}$. In NED coordinates, the $x_n$ axis points to true North, the $y_n$ axis points to the East and the $z_n$ axis points downwards, normal to Earth's surface. The NED-frame is considered to be inertial for local navigation, so that Newton's laws of motion still apply. In the body-frame, the $x_b$ axis points along the longitudinal axis of the vessel, the $y_b$ axis points along the transverse axis and the $z_b$ axis is normal to the surface of the vessel. \autoref{fig:ref_frames} illustrates the relationship between the two reference frames.
        The rotation of \{b\} with respect to \{n\} is described by the Euler angle rotation matrix: 
        \begin{equation} \label{eq:rotMatrix}
        \resizebox{0.8\linewidth}{!}{%
            $\mathbf{R}_b^n (\boldsymbol{\Theta}_{nb}) = 
            \begin{bmatrix}
            c\psi c\theta & - s\psi c\phi + c\psi s\theta s\phi & s\psi s\phi + c\psi c\phi s\theta \\
            
            s\psi c\theta & c\psi c\phi + s\phi s\theta s\psi & -c\psi s\phi + s\theta s\psi c\phi \\
            
            - s\theta & c\theta s\phi & c\theta c\phi
            \end{bmatrix}$%
            }
        \end{equation}
        
        \noindent where $\boldsymbol{\Theta}_{nb} = [\phi, \theta, \psi]$ are the Euler angles describing the vehicle's attitude, and $s\phi = \sin{\phi}$, $c\phi = \cos{\phi}$. Translating vectors from \{b\} to \{n\} is done by multiplying by the rotation matrix. The inverse rotation, i.e. from \{n\} to \{b\}, is given by $R_n^b = (R_b^n)^T$. 
        
        The kinematic state vector is the concatenation of the position of the vehicle in NED coordinates and the Euler angles, symbolized by $\boldsymbol{\eta} = [\mathbf{p}^n, \boldsymbol{\Theta}_{nb}]^T = [x, y, z, \phi, \theta, \psi]^T$. The differential equation describing the position of the AUV is found buy rotating the well defined velocity vector in \{b\}, $\mathbf{v}^b= [u, v, w]^T$, to the NED frame:
        \begin{equation}
            \label{eq:pdiff}
            \Dot{\mathbf{p}}^n = \mathbf{R}_b^n(\boldsymbol{\Theta}_{nb})\mathbf{v}^b
        \end{equation}
        
        Similarly, we transform the angular velocities in \{b\}, $\boldsymbol{\omega}_{n/b}^b$, to \{n\} with the transformation matrix $\mathbf{T}_\Theta (\boldsymbol{\Theta}_{nb})$:
        
        \begin{equation}
            \label{eq:eulerdiff}
            \Dot{\boldsymbol{\Theta}}_{nb} = \mathbf{T}_\Theta (\boldsymbol{\Theta}_{nb}) \boldsymbol{\omega}_{b/n}^b
            = 
            \begin{bmatrix}
            1 & s\phi t\theta & c\phi t\theta \\
            0 & c\phi & -s\phi \\
            0 & \frac{s\phi}{c\theta} & \frac{c\phi}{c\theta}
            \end{bmatrix}
            \begin{bmatrix}
            q \\
            p \\
            r 
            \end{bmatrix} 
        \end{equation}
        where $t\theta = \tan{\theta}$. \footnote{Remark: This transformation is not well-defined for $\theta = \frac{\pi}{2}$, which corresponds to an attitude where the AUV's x-axis is parallel to the $z_n$-axis in \{n\}. This singularity can be avoided by applying quaternion parameterization~\cite[p~.27]{Fossen}. However, for our application, this attitude is considered as being outside the AUV's area of operation and thus, the Euler angles are used.}   
        Combining \autoref{eq:pdiff} and \autoref{eq:eulerdiff} yields the full kinematic differential equation:
        \begin{equation} \label{}
            \Dot{\boldsymbol{\eta}} = 
            \begin{bmatrix}
            \Dot{\mathbf{p}}^n \\
            \Dot{\boldsymbol{\Theta}}_{nb}
            \end{bmatrix} =
            \begin{bmatrix}
            \mathbf{R}_b^n(\boldsymbol{\Theta}_{nb}) & \mathbf{0} \\
            \mathbf{0} & \mathbf{T}_{\Theta}(\boldsymbol{\Theta}_{nb})
            \end{bmatrix} 
            \begin{bmatrix}
            \mathbf{v}^b \\
            \boldsymbol{\omega}_{b/n}^b    
            \end{bmatrix}=
            \mathbf{J}_{\Theta}(\boldsymbol{\eta}) \boldsymbol{\nu}
        \end{equation}
        The Kinetic equations of motion for a marine craft can essentially be distilled into a mass-spring-damper system. The spring forces are the restoring forces acting on the body due to buoyancy, while the damping is a result of the hydrodynamic forces caused by motion. The model presented here is based on the work of da Silva et. al~\cite{AUV-sim}, and is valid under the assumptions that
        
        \begin{itemize}
            \item the AUV operates at a depth where disturbances from wind and waves are negligible
            \item the maximum expected speed is $2m/s$
            \item mass is distributed such that the moments of inertia can be approximated by those of a spheroid
            \item the center of gravity (COG) is located $z_G=1cm$ under the center of buoyancy (COB) to create a restoring moment in roll and pitch
            \item the AUV is top-bottom and port-starboard symmetric
            \item the AUV is slightly buoyant, as a fail-safe mode in case of power loss
        \end{itemize}
        
        Moreover, the model's numerical values are based on the specifications given in \autoref{tab:AUVspec}.
        
        \begin{table}[pos=h]
        \centering
        \caption{Specifications for the simulated AUV \cite{AUV-sim}}
        \begin{tabular}{||c|l|c|c|| }
        \hline
        \textbf{Symbol} & \textbf{Description} & \textbf{Value} & \textbf{Unit} \\
        \hline
        $m$ & Mass & 18 & $kg$ \\
        $L$ & Length & 108 & $cm$ \\
        $W$ & Weight & 176 & $N$ \\
        $B$ & Buoyancy & 177 & $N$ \\
        $z_G$ & COG w.r.t. COB in z-axis & 1 & $cm$ \\
        $d$ & Diameter & 15 & $cm$ \\
        \hline
        \end{tabular}
        \label{tab:AUVspec}
        \end{table}
        
        The vessel's motion in \{b\}is governed by the following nonlinear kinetic equations:
        
        \begin{equation}\label{eq:eom}
        \resizebox{0.85\linewidth}{!}{%
            $\underbrace{\mathbf{M}\Dot{\boldsymbol{\nu}}_r}_{\mbox{\footnotesize Mass forces}} +
            \underbrace{\mathbf{C}(\boldsymbol{\nu}_r)\boldsymbol{\nu}_r}_{\mbox{\footnotesize Coriolis forces}} + 
            \underbrace{\mathbf{D}(\boldsymbol{\nu}_r)\boldsymbol{\nu}_r}_{\mbox{\footnotesize Damping forces}} +
            \underbrace{\mathbf{g(\boldsymbol{\eta}})}_{\mbox{\footnotesize Restoring forces}} = 
            \boldsymbol{\tau}_{control}$%
            }
        \end{equation}
        where $\boldsymbol{\nu}_r$ is the velocity relative to the velocity of ocean currents. Initially it is assumed that there are no currents. However, as part of the motivation behind this work is to uncover how the MLC handles the inclusion of environmental disturbances, the model is implemented such that ocean currents can easily be added.
        
        \subsubsection{Mass Forces}
            The system's inertia matrix, $\mathbf{M}$, is the sum of the inertia matrix for the rigid body and the added mass. Added mass is the inertia added from the weight of fluid displaced by the vessel when moving through it. Because of the symmetry assumptions, both matrices are diagonal. However, the rigid body matrix is defined in the center of gravity, such that it must be shifted to the center of control, yielding some coupling terms:
            \begin{equation}
            \begin{gathered}
                \mathbf{M} =
                \left[
                \begin{matrix}
                m - X_{\Dot{u}} & 0 & 0 & 0\\
                0 & m - Y_{\Dot{v}} & 0 & -m z_G \\
                0 & 0 & m - Z_{\Dot{w}}  & 0 \\
                0 & -m z_G & 0 & I_x - K_{\Dot{p}}\\
                m z_G & 0 & 0 & 0  \\
                0 & 0 & 0  & 0
                \end{matrix}
                \right.
                \\
                \left.
                \begin{matrix}
                m z_G & 0 \\
                 0 & 0 \\
                 0 & 0 \\
                0 & 0 \\
                 I_y - M_{\Dot{q}} & 0 \\
                0 & I_z - N_{\Dot{r}} 
                \end{matrix}
                \right]
            \end{gathered}
            \end{equation}
        
        \subsubsection{Coriolis Forces}
            Naturally, the added mass will also effect the Coriolis-centripetal matrix, $\mathbf{C(\boldsymbol{\nu}_r})$, which defines the forces occurring due to \{b\} rotating about \{n\}. Moreover, the Coriolis-centripetal matrix could be expressed independently of the linear velocities, easing the implementation of irrotational ocean currents~\cite[p.~222]{Fossen}{}. In 6 DOF, these matrices are given by:
            
            \begin{equation}
                \mathbf{C}(\boldsymbol{\nu}_r) =
                \begin{bmatrix}
                \mathbf{0} && \mathbf{C}_{12}(\boldsymbol{\nu}_r)
                \\
                \mathbf{C}_{21}(\boldsymbol{\nu}_r) && \mathbf{C}_{22}(\boldsymbol{\nu}_r)
                \end{bmatrix} \\
            \end{equation}
            
            \noindent where
            \begin{equation}
            \begin{gathered}
                \mathbf{C}_{12}(\boldsymbol{\nu}_r) = 
                \left [
                \begin{matrix}
                m z_G r &  (m-Z_{\Dot{w}})w_r \\
                - (m-Z_{\Dot{w}})w_r &  m z_G r  \\
                -m z_G p + (m-Y_{\Dot{v}})v_r & 
                -m z_G q - (m-X_{\Dot{x}})u_r
                \end{matrix}
                \right.
                \\
                \left.
                \begin{matrix}
                -(m-Y_{\Dot{v}})v_r \\
                (m-X_{\Dot{u}})u_r \\
                0
                \end{matrix}
                \right ]
                \\
                \mathbf{C}_{21}(\boldsymbol{\nu}_r) = 
                \left [
                \begin{matrix}
                -m z_G r &  (m-Z_{\Dot{w}})w_r \\
                - (m-Z_{\Dot{w}})w_r & -m z_G r  \\
                (m-Y_{\Dot{v}})v_r & 
                - (m-X_{\Dot{u}})u_r 
                \end{matrix}
                    \right.
                \\
                \left.
                \begin{matrix}
                m z_G p - (m-Y_{\Dot{v}})v_r \\
                m z_G q + (m-X_{\Dot{x}})u_r \\
                0
                \end{matrix}
                \right ]
                \\
                \mathbf{C}_{22}(\boldsymbol{\nu}_r) =
                \left [
                \begin{matrix}
                0 &  (I_z-N_{\Dot{r}})r \\
                -(I_z-N_{\Dot{r}})r &  0 \\
                (I_y-M_{\Dot{q}})q & 
                -(I_x-K_{\Dot{p}})p 
                \end{matrix}
                        \right.
                \\
                \left.
                \begin{matrix}
                -(I_y-M_{\Dot{q}})q \\
                (I_x-K_{\Dot{p}})p \\
                0
                \end{matrix}
                \right ]
            \end{gathered}
            \end{equation}
            Combining and inserting numerical values yields the full Coriolis-centripetal matrix: 
            
            \begin{equation}
            \begin{gathered}
                \mathbf{C}(\boldsymbol{\nu}_r) =
                \left [
                \begin{matrix}
                0 & 0 & 0 & 0.18r  \\
                0 & 0 & 0 & -34w_r \\
                0 & 0 & 0 & -0.18p + 34v_r \\
                -0.18r & 34w_r & 0.18p - 34v_r & 0  \\
                -34w_r & -0.18r & 0.18q + 19u_r & -1.8r \\
                34v_r & -19u_r & 0 & 1.8q
                \end{matrix}
                \right.
                \\
                \left.
                \begin{matrix}
                34w_r & -34v_r \\
                0.18r & 19u_r \\
                -0.18q - 19u_r & 0 \\
                1.8r & -1.8q \\
                0 & 0.04p \\
                -0.04p & 0
                \end{matrix}
                \right ]
                \end{gathered}
            \end{equation}
            
        \subsubsection{Damping Forces}
            The components of hydrodynamic damping modelled is linear viscous damping, nonlinear (quadratic) damping due to vortex shedding and lift forces from the body and control fins. Thus, the damping matrix, $\mathbf{D}(\boldsymbol{\nu}_r)$, can be expressed as:
            
            \begin{equation}
                \mathbf{D}(\boldsymbol{\nu}_r) = \mathbf{D} + \mathbf{D}_n(\boldsymbol{\nu}_r) + \mathbf{L}(\boldsymbol{\nu}_r)
            \end{equation} 
            
            \noindent The linear damping is given by 
            
            \begin{equation*}
            \begin{gathered}
                \mathbf{D} =
                - \begin{bmatrix}
                X_u & 0 & 0 & 0 & 0 & 0 \\
                0 & Y_v & 0 & 0 & 0 & Y_r \\
                0 & 0 & Z_w & 0 & Z_q & 0 \\
                0 & 0 & 0 & K_p & 0 & 0 \\
                0 & 0 & M_w & 0 & M_q & 0 \\
                0 & N_v & 0 & 0 & 0 & N_r
                \end{bmatrix}
            \end{gathered}
            \end{equation*}
            The nonlinear damping is given by
            \begin{equation}
            \begin{gathered}
                \mathbf{D}_n({\boldsymbol{\nu}_r})= -
                \left [
                \begin{matrix}
                X_{u|u|}|u| & 0 & 0 & 0 \\
                0 & X_{v|v|}|v| & 0 & 0 \\
                0 & 0 & Z_{w|w|}|w| & 0 \\
                0 & 0 & 0 & K_{p|p|}|p| \\
                0 & 0 & M_{w|w|}|w| & 0  \\
                0 & N_{v|v|}|v| & 0 & 0 
                \end{matrix} 
                \right.
            \\
            \left.
                \begin{matrix}
                 0 & 0 \\
                 0 & Y_{r|r|}|r| \\
                 Z_{q|q|}|q| & 0 \\
                 0 & 0 \\
                 M_{q|q|}|q| & 0 \\
                 0 & N_{r|r|}|r|
                \end{matrix}
                \right ]
            \end{gathered}
            \end{equation}
            
            \noindent Finally, the lift is given by
            \begin{equation}
            \begin{gathered}
                \mathbf{L}({\boldsymbol{\nu}_r})= - 
                \left [
                \begin{matrix}
                0 & 0 & 0 & 0 \\
                0 & Y_{uv_f} + Y_{uv_b} & 0 & 0 \\
                0 & 0 & Z_{uw_f} + Z_{uw_b} & 0\\
                0 & 0 & 0 & 0 \\
                0 & 0 & M_{uw_f} + M_{uw_b} & 0 \\
                0 & N_{uv_f} + N_{uv_b} & 0 & 0
                \end{matrix} 
                \right.
                \\
                \left.
                \begin{matrix}
                0 & 0 \\
                0 & Y_{ur_f} \\
                Z_{uq_f} & 0 \\
                0 & 0 \\
                M_{uq_f} & 0 \\
                0 & N_{ur_f}
                \end{matrix} 
                \right ]u_r
            \end{gathered}
            \end{equation}
            
        \subsubsection{Restoring Forces}
            Buoyancy acting on the body initiates restoring forces and moments for the AUV. This can be considered as a virtual spring acting on the system. Based on all previous assumptions, the restoring force vector can be written as:
            \begin{equation}
                \mathbf{G}(\boldsymbol{\eta}) = 
                \begin{bmatrix}
                (W-B) \sin{\theta} \\
                -(W-B) \cos{\theta} \sin{\phi} \\
                -(W-B) \cos{\theta} \cos{\phi} \\ 
                z_G W \cos{\theta} \sin{\phi} \\
                z_G W \sin{\theta} \\
                0
                \end{bmatrix}
            \end{equation}
            \subsubsection{Control Inputs}
            There are 3 control inputs: propeller shaft speed and rudder and elevator fin rotation denoted by $n$, $\delta_r$ and $\delta_s$, respectively. The control surfaces can maximally be rotated \ang{30} in each direction, and the propeller thrust is limited such that the AUV does not violate the low-speed assumption. The control inputs are related to the control force vector according to \autoref{eq:input}:
            \begin{equation}
                \boldsymbol{\tau}_{control} =  \begin{bmatrix}
                1 & 0 & 0 \\
                0 & Y_{uu\delta_r} u_r^2 & 0 \\
                0 & 0 & Z_{uu\delta_s} u_r^2 \\
                0 & 0 & 0 \\
                0 & 0 & M_{uu\delta_s} u_r^2 \\
                0 & N_{uu\delta_r} u_r^2 & 0 \\
                \end{bmatrix}
                \begin{bmatrix}
                n \\
                \delta_r \\
                \delta_s 
                \end{bmatrix}
                \label{eq:input}
            \end{equation}
            
            For a more thorough derivation of the model and how the numerical values are calculated, the readers are referred to \cite{AUV-sim} and \cite{Fossen}. 
    \subsection{Path-following}\label{sec:PF}
        The main control problem addressed in this work is that of path-following, where the objective is to follow a pre-planned path without time-constraints. Consequently, the goal is to drive tracking-errors to zero~\cite[ch.~9]{Fossen}. A set of $n$ waypoints is used to represent the path, starting at the origin of the NED coordinate frame for simplicity. The path is generated by linear interpolation between the waypoints, resulting in a piecewise straight-line path. For a path defined in three-dimensional space, the parametric equations for the interpolation scheme are~\cite{AUV-guidance}:
        \begin{equation}
        \begin{gathered}
            x_{p,i}(s) = x_{p,i-1} + s \cos{\chi_{p,i-1}} \cos{\upsilon_{p,i-1}} \\
            y_{p,i}(s) = y_{p,i-1} + s \sin{\chi_{p,i-1}} \cos{\upsilon_{p,i-1}} \\
            z_{p,i}(s) = z_{p,i-1} - s \sin{\upsilon_{p,i-1}}
        \end{gathered}
        \end{equation}
        where subscript $p$ signifies that the coordinate is representing the path and $i$ denotes the waypoint index. These coordinates define the path relative to the inertial frame. The angles $\chi_{p,i-1}$ and $\upsilon_{p,i-1}$ denote the azimuth and elevation angle of the straight line between waypoints $i-1$ and $i$ relative to \{n\}. The parametric equations are continuously differentiable with respect to $s$, which is the along-track distance travelled on the path from waypoint $i-1$ to $i$.
        
\begin{figure}[pos=h]
            \centering
            \includegraphics[width=\linewidth]{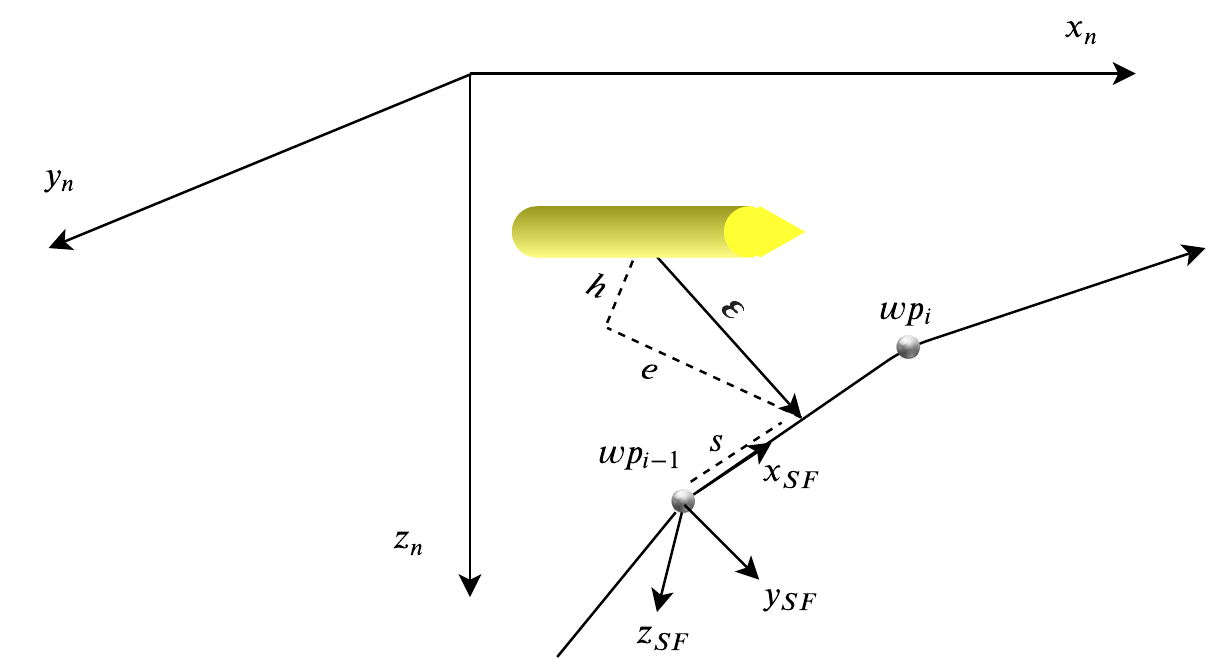}
            \caption{The Serret-Frenet reference frame defines the components for the tracking-error vector. The control objective in path-following is to drive $e$ and $h$ to zero.}
            \label{fig:SFref}
        \end{figure}

        To define the tracking-errors, the Serret-Frenet (\{SF\}) reference frame associated with each point of the path is introduced. The $x_{SF}$ axis is tangent to the path, the $y_{SF}$ axis normal to the path and the $z_{SF}$ axis is given by $z_{SF} = x_{SF} \times y_{SF}$, and is thus orthogonal to the other two axes~\cite{encarnacao20003d}. The vector $\boldsymbol{\varepsilon} = [s, e, h]^T$ is defined by the along-track distance, cross-track error and vertical-track error illustrated in \autoref{fig:SFref}. This vector points towards the closest point on the path from the vessel.
        
        Traditional path-following controllers strive to align the velocity vector in the inertial frame with the path tangent. Instead of aiming directly at the closest point of the path, the vessel aims at a point further ahead, decided by a look-ahead distance $\Delta$, which is set by the control designer.\footnote{In general the look-ahead point is on the path tangent, i.e. on $X_{SF}$, but as straight-line paths are considered, a path-segment and its tangent align.} The vector $\boldsymbol{\varepsilon}$ is obtained in the {SF}-frame by:
        
        \begin{equation}
            \boldsymbol{\varepsilon} = \mathbf{R}_n^{SF}(\upsilon_p, \chi_p)(\mathbf{p}^n-\mathbf{p}_p^n)
        \end{equation}
        
        \noindent where $\mathbf{p}$ is the position of the vessel and $\mathbf{p}_p$ is the closest point on the path. $\mathbf{R}_n^{SF}(\upsilon_p, \chi_p)$ is the rotation matrix from \{n\} to \{SF\}, which has the same form as the rotation matrix in \eqref{eq:rotMatrix} with $\phi=0$, $\theta=\upsilon_p$ and $\psi = \chi_p$. Now the desired azimuth and elevation angle can be calculated according to:
        
        \begin{equation}
            \chi_d(e) = \chi_p + \chi_r(e) \;\;\;\;\;,\;\;\;\;\; \upsilon_{d}(h) = \upsilon_p + \upsilon_r(h)
            \label{eq:des_ang}
        \end{equation}
        \noindent where
        \begin{equation}
            \chi_r(e) = \arctan(-\frac{e}{\Delta}),\quad
            \upsilon_r(h) = \arctan(\frac{h}{\sqrt{e^2 + \Delta^2}})
            \label{eq:cor_ang}
        \end{equation}
        The angles, $\chi_r(e)$ and $\upsilon_r(h)$, can be interpreted as corrective steering, and when driven to zero the velocity vector aligns perfectly with the tangent of the path~\cite{AUV-guidance}. This provides a reasonable control objective, and the challenge now lies in finding a control law that maps these errors to good actuator outputs.
        
        A few approaches can be used to design these control laws. One is to decompose the system into a longitudinal and a lateral model and design autopilots for each of them. The coupling terms are considered disturbances. If the motion is reserved for one plane at a time, these disturbances will remain rather small and neglecting them is justified. Additionally, the system can be linearized about an equilibrium point and pole-placement or optimal control strategies can be used to derive the feedback gains~\cite[ch.~10]{Fossen}.
        
        Encarnasi and Pascoal proposed a more ambitious approach (\cite{encarnacao20003d}) in which, they developed a nonlinear kinematic controller using Lyapunov theory, feedback linearization and backstepping. Their simulations showed impressive results for 3D path-following, both for straight-line paths and a helix. However, disturbances and saturation limits for the actuators were not considered in the analysis.
        
    \subsection{Deep Reinforcement Learning}
        The DRL techniques were developed for learning control systems as early as 1965~\cite{OldLearning}. Here we give a brief description of the relevant theory behind the technique.
        
        \subsubsection{\textbf{Reinforcement Learning}}
        In RL an algorithm, known as an \textbf{agent}, makes an \textbf{observation} $s_t$ of an \textbf{environment} and performs an \textbf{action} $a_t$. The observation is referred to as the state of the system, and is drawn from the state space $\mathcal{S}$. The action is restricted to the well-defined action space $\mathcal{A}$. When an RL task is not infinitely long, but ends at some time $T$, we say that the problem is episodic, and that each iteration through the task is an episode.
        
        After performing an action, the agent receives a scalar \textbf{reward} signal $r_t=r(s_t,a_t)$. The reward quantifies how good it was to choose action $a_t$ when in state $s_t$. The objective of the agent is typically to maximize expected cumulative reward.
        
        The action choices of the agent are guided by a \textbf{policy} $\pi(s)$, which can be either deterministic or stochastic. In the case that the learning algorithm involves a neural network, the policy is parametrized by the learnable parameters of the network, denoted by $\theta$. When the policy is stochastic and dependent on a neural network, we write $\pi(s) = \pi_{\theta}(a|s)$.
        
        The \textbf{value function} $V_{\pi}(s)$ describes how valuable it is to be in state $s$ under the policy $\pi$. The action-value funtcion, or \textbf{Q-function} $Q_\pi(s,a)$ expresses the value of performing action $a$ when in state $s$.
        
        Mathematically, the value function and Q-function can be expressed through the \textbf{return}, which is shown in \autoref{eq:return}. The return is a measure of accumulated reward between times $t$ and $T$, where $T$ might be infinity. The discount factor $\gamma$ weights the importance of rewards that are close in time versus those distant in time.
        
        \begin{equation}\label{eq:return}
            R_t^\gamma = \sum_{k=t}^T \gamma^{t-k}r(s_k,a_k), \: 0 < \gamma < 1
        \end{equation}
        
        Using the return, the value function is $V_{\pi}(s_t) = \mathbb{E}\{R_t^\gamma|s_t;\pi\}$ and the Q-function is $Q_{\pi}(s_t,a_t) = \mathbb{E}\{R_t^\gamma|s_t,a_t;\pi\}$.
        
        Learning by reinforcing good choices is synonymous with how humans (and other animals) learn. RL is therefore a formal version of trial-and-error learning. The goal of many RL algorithms can be formally stated as the optimization problem in \autoref{eq:RLobjective}~\cite{Sutton}.
        
        \begin{equation}\label{eq:RLobjective}
            \theta^* = \text{arg} \max_\theta \mathbb{E}_{s \sim \rho_{\theta}, a \sim \pi_{\theta}} \left[R_t^\gamma\right]
        \end{equation}
        
\begin{figure}[pos=h]
            \centering
            \includegraphics[width = 0.85\linewidth]{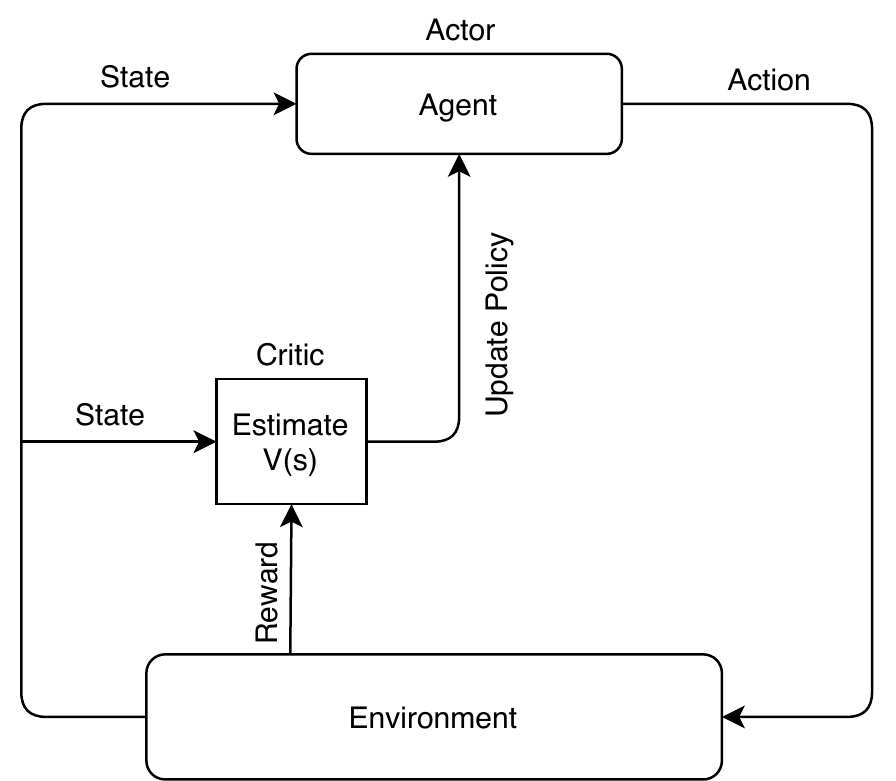}
            \caption{Actor-critic methods are a hybrid of policy gradient and value-based methods.}
            \label{fig:A2C}
        \end{figure}

        Solving \autoref{eq:RLobjective} yields the optimal parameters $\theta=\theta^*$ that maximize the expected return at all times $t$, when the actions are drawn from the policy $\pi_{\theta}$, and the state distribution is given by $\rho_{\theta}$. Algorithms that aim to solve \autoref{eq:RLobjective} can be roughly divided into four categories: 
        
        \begin{itemize}
            \item \textbf{Policy gradients method}: Maximize the objective directly through gradient ascent~\cite{PolGrad}.
            \item \textbf{Value-based methods}: Estimate the value function and/or the Q-function, and make a policy that increases the probability of taking actions that maximize their values \cite{Sutton}.
            \item \textbf{Actor-Critic methods}: A hybrid of policy gradient and value-based methods. The value function or Q-function is approximated by a neural network which acts as a critic. The actor's policy is updated as suggested by the critic through some policy gradient method \cite{A2C}. This idea is illustrated in \autoref{fig:A2C}.
            \item \textbf{Model-based RL}: A model of the environment is created through exploration, and the estimate is utilized to make decisions. For instance, the model can be used in optimal control or value-based methods \cite{ModRL}.
        \end{itemize}
        
        \subsubsection{\textbf{Proximal Policy Optimization}}
        In this work, we use the actor-critic algorithm known as Proximal Policy Optimization (PPO) as suggested by Schulman et al. \cite{PPO-article}. In this section, an overview of the theory is given. The advantage function in the PPO is defined as:
        \begin{equation}
            A(s,a) = Q(s,a) - V(s).
        \end{equation}
        \noindent The advantage function represents the difference in expected return by taking action $a$ in state $s$, as opposed to following the policy. Because both $Q(s,a)$ and $V(s)$ are unknown, an estimate of the advantage function, $\hat{A_t}$, is calculated based on an estimate of the value function $\hat{V}(s)$, which is made by the critic neural network. 
        
        A method for estimating the advantage function is the generalized advantage estimate (GAE), given in \autoref{eq:gae}~\cite{schulman2015highdimensional}.
        
        \begin{equation}\label{eq:gae}
        \begin{gathered}
            \hat{A}_t = \delta_t + (\gamma \lambda) \delta_{t+1} + \dots + (\gamma \lambda)^{T-t+1} \delta_{T-1} \\
            \text{where } \delta_t = r_t + \gamma \hat{V}(s_{t+1}) - \hat{V}(s_t)
        \end{gathered}
        \end{equation}
        
        Here, $T$ is a truncation point which is typically much smaller than the duration of an entire episode. As before, $\gamma$ is the discount factor. As the GAE is a sum of uncertain terms, the tuneable parameter $0 \leq \lambda \leq 1$ is introduced to reduce variance. However, $\lambda < 1$ makes the GAE biased towards the earlier estimates of the advantage function. Hence, choosing $\lambda$ is a bias-variance trade-off.
        
        The second key component in PPO is introducing a surrogate objective. It is hard to apply gradient ascent directly to the RL objective in \autoref{eq:RLobjective}. Therefore, Schulman et al. suggest a surrogate objective which is such that an increase in the surrogate provably leads to an increase in the original objective ~\cite{PPO-article}. The proposed surrogate objective function is given by \autoref{eq:PPOobjective}.
        
        \begin{equation}\label{eq:PPOobjective}
        \resizebox{0.85\linewidth}{!}{%
            $L^{CLIP}(\theta) = \hat{\mathbb{E}}_t \left[ \min \left( \frac{\pi_{\theta}(a_t | s_t)}{\pi_{\theta_{old}}(a_t | s_t)}\hat{A}_t,  clip \left( \frac{\pi_{\theta}(a_t | s_t)}{\pi_{\theta_{old}}(a_t | s_t)},1-\epsilon, 1+\epsilon \right) \hat{A}_t \right) \right]$%
            }
        \end{equation}
        
        The tuning parameter $\epsilon$ reduces the incentive to make very large changes to the policy at every step of the gradient ascent. This is necessary as the surrogate objective only estimates the original objective locally in a so-called trust-region.

\section{Implementation}
\label{section:implementation}
    The implementation of our solution makes use of the RL framework \textbf{OpenAI Gym}. OpenAI Gym is a Python library which was created for the purpose of standardizing the benchmarks used in RL research~\cite{openai-gym}. It provides an easy-to-use framework for creating RL environments in which custom RL agents can be deployed and trained with minimal overhead. Stable Baselines is a Python library that provides a large set of state-of-the-art parallelizable RL algorithms compatible with the OpenAI gym framework, including PPO~\cite{stable-baselines}. PPO is used in this work because of its reputable performance on continuous control problems and stability \cite{meyer2020taa}. In fact, its performance on the OpenAI benchmark - a set of standardized test environments, created to assess and compare different RL algorithms - was so impressive that it has become the go-to RL algorithm in the OpenAI library. The  algorithm in its most general form can be seen in Algorithm \ref{alg:PPO}.
    
    \begin{algorithm}
    \caption{Proximal Policy Optimization, Actor-Critic style}
    \SetAlgoLined
    \For{iteration: 1,2...}{
    \For{actor: 1,2...N}{
    \text{Run policy $\pi_{\theta_{old}}$ for T time-steps} \\
    \text{Compute advantage estimates $\hat{A}_1...\hat{A}_T$}
    }
    \text{(Optimize surrogate L wrt $\theta$, for K epochs} \\  
    \text{with mini-batch size  $M<NT$)} \\
    \textbf{$\theta_{old} \leftarrow \theta$}
    }
    \label{alg:PPO}
    \end{algorithm}
    More details of the implementation can be found in the code on Github~\cite{github-code}. 

\section{Simulation set-up}
\label{section:experiments}
    This section provides a detailed description of the set up used to do path following simulations presented in the subsequent sections. We utilized two distinct approaches to train the RL agent to achieve the objective of path following: one is called end-to-end training (Section \ref{subsec:endtoend}) where the agent learns through freely exploring the environment and gradually trading off exploration for exploitation. In the second approach, the agent's learning is assisted by PID controllers (Section \ref{subsec:pid}). Since one of the desired features of the path-following problem is to maintain a desired cruise speed while minimizing tracking-errors, the first part of the controller design was a simple velocity controller. Because of the reduced complexity of this problem, it could act as a sanity check for the simulator code and the implementation of the RL scheme. Furthermore, because a constant reference cruise speed is part of both the approaches, the reward functions should contain similar penalizing terms for deviating from this reference. The velocity controller worked well but to save space we do not present the result from those simulations. In the two approaches presented here, the AUV is randomly initialized within a proximity of 5 meters of the first waypoint. It is desired that the AUV maintains a cruise speed of $u_d = 1.5~ms^{-1}$. The AUV is underactuated, as it operates in 6 DOF with only 3 actuators. The simulated learning processes are described in the following sections.
    
    \subsection{End-to-end learning}
    \label{subsec:endtoend}
\begin{figure}[pos=h]
            \centering
            \includegraphics[width=1\linewidth]{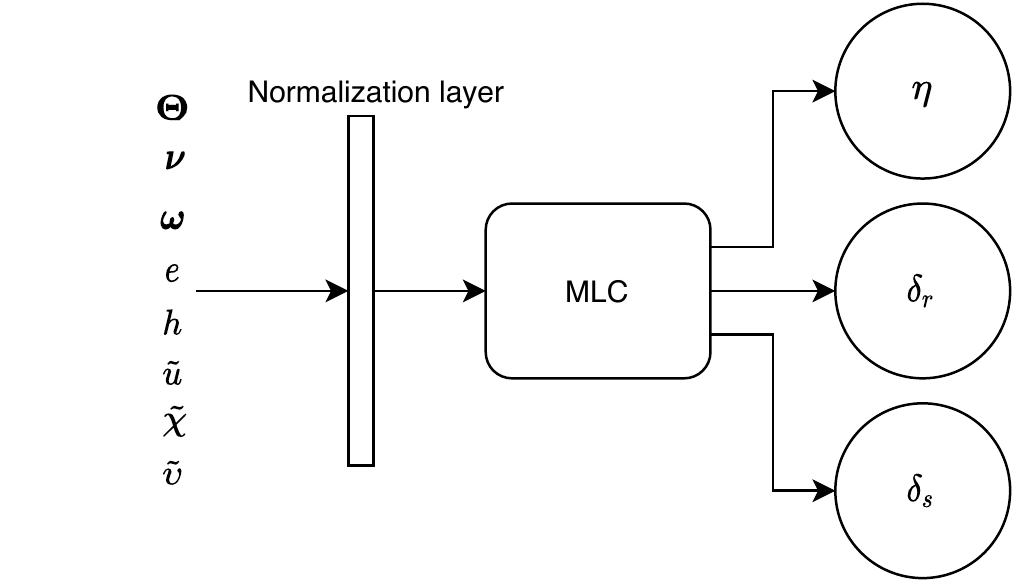}
            \caption{End-to-end modeling approach}
            \label{fig:MLC}
        \end{figure}

        In this scenario (Figure \ref{fig:MLC}), the observation that the agent makes of the environment, $s$, consists of normalized measurements of the Euler angles, $\boldsymbol{\Theta}_o = [\phi_o, \theta_o, \psi_o]^T$, the angular rates $\boldsymbol{\omega}_o = [p_o, q_o, r_o]^T$ and the control errors $\boldsymbol{\epsilon} = [\tilde{u}_o, \tilde{\chi}_o, \tilde{\upsilon}_o, e_o, h_o]^T$. These are listed in Table \ref{tab:PF_obs} together with their empirical or true maximums, and reward function coefficients. Subscript $_o$ indicates that these values have been normalized by their empirical or true maxima, to ensure that values fed to the neural networks are between \num{-1} and \num{1}. Neural networks work better with normalized data, which often improves the numerical stability of the model and reduces training time. There is no obvious empirical maximum for the vertical and cross track error, which make the normalization factors a design choice. Choosing $e_{max} = h_{max} = 25m$ is reasonable, since tracking-errors $>25m$ indicate significantly poor performance. Errors above this threshold can therefore be uniform. As is the state-of-the art in RL and ML in general, the reward function was crafted through reasoning and iteratively modified. The reward function governing the AUV behaviour is 
        \begin{equation}
            \label{eq:PF_reward}
            r_t = \boldsymbol{\alpha}_1^T |\boldsymbol{\Theta}_o| + \boldsymbol{\alpha}_2^T |\boldsymbol{\omega_o}| + \boldsymbol{\alpha}_3^T |\boldsymbol{\epsilon}|.
        \end{equation}
        The penalization factors, $\boldsymbol{\alpha}_i$, were chosen based on the control objective and empirical trials. For instance, it is not obvious that roll and roll rate should be penalized, but this is done to avoid a behaviour where the rudder acts as the elevator fin and vise versa. Angular rates are penalized to indirectly penalize aggressive and large control inputs. In \autoref{eq:PF_reward}, $|\cdot|$ denotes the element-wise absolute value. As all elements of $\boldsymbol{\alpha}_i$ are negative, the reward function is always negative. The chosen penalization factors $\boldsymbol{\alpha}_i$ are given in Table \ref{tab:PF_obs}.
            
        \begin{table}[pos=h]
        \centering
        \caption{Observations the agent inputs when doing path following. The notation ${\tilde{\cdot}}$ symbolizes the difference between the desired and the actual value, in accordance with the standard control literature notation.}
        \renewcommand*{\arraystretch}{1.5}
        \begin{tabular}{l c c c}
        \hline
        \textbf{Observation} & & \textbf{Max} & \textbf{$\alpha$} \\
        \hline
        Roll & $\phi_o = \frac{\phi}{\phi_{max}} \in [-1,1]$ & $\pi$ & \num{-5e-3} \\
        Pitch & $\theta_o = \frac{\theta}{\theta_{max}} \in [-1,1]$ & $\pi$ & \num{0} \\
        Yaw & $\psi_o = \frac{\psi}{\psi_{max}} \in [-1,1]$ & $\pi$ & \num{0}\\
        Roll rate & $p_o =  \frac{p}{p_{max}} \in [-1,1]$ & \num{1.2} & \num{-5e-3}\\
        Pitch rate & $q_o = \frac{q}{q_{max}} \in [-1,1]$ & \num{0.4} & \num{-5e-4}\\
        Yaw rate & $r_o = \frac{r}{r_{max}} \in [-1,1]$ & \num{0.4} &\num{-5e-4}\\
        Surge error & $\tilde{u}_o = \frac{u_d-u}{u_{max}} \in [-1,1]$ & \num{2} & \num{-5e-3} \\
        Course error & $\tilde{\chi}_o = \frac{\chi_d - \chi}{\chi_{max}} \in [-1,1]$ & $\pi$ & \num{-2.5e-3}\\
        Elevation error & $\tilde{\upsilon}_o = \frac{\upsilon_d - \upsilon}{\upsilon_{max}} \in [-1,1]$ & $\pi$ & \num{-2.5e-3} \\
        Cross track error & $e_o = \frac{e}{e_{max}} \in [-1,1]$ & \num{25} & \num{-5e-3}\\
        Vertical track error & $h_o = \frac{h}{h_{max}} \in [-1,1]$ & \num{25} & \num{-5e-3} \\
        \end{tabular}
        \label{tab:PF_obs}
        \end{table}
    \subsection{PID assisted learning}
    \label{subsec:pid}
\begin{figure}[pos=h]
            \centering
            \includegraphics[width=1\linewidth]{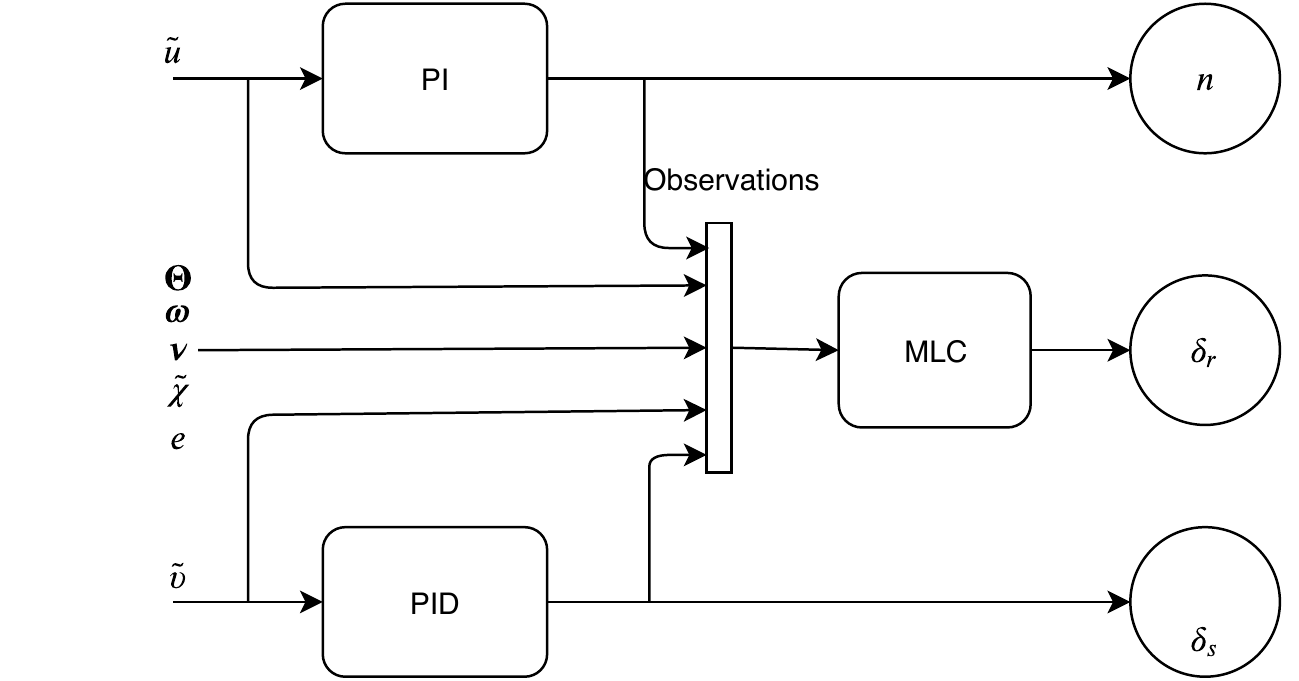}
            \caption{A PID assisted approach to MLC training, as an attempt to improve the speed of learning and avoid the unlearning phenomenon discussed in \hyperref[section:results]{Section 5}. The agent observes all information in the system. Because the surge speed can be modeled as a simple first order system, no derivative action is included in this controller. Hence, this is a PI-controller.}
            \label{fig:PID_MLC}
        \end{figure}

        In PID assisted training, the idea is to decouple the neural network into three parts that will be trained separately, each controlling its own actuator. A cross-track controller operates the rudder fins, a vertical-track controller the elevator fins, and a velocity controller controls the propeller shaft speed. This is analogous to how traditional autopilots are designed, but the key difference is in the set-up: To not lose information about the system it should see every state and other actuator while training and in operation. In order to avoid training the neural networks together, PI/PID controllers are enabled to stabilize the two sub-processes that are not considered at the time. An example of this scheme when training the network controlling the rudder is illustrated in \autoref{fig:PID_MLC}. The design of the PID controllers is not central, but the PID controllers should not lead to unstable behaviour. One could argue that a controller that does not make the AUV behave perfectly is preferable to an optimal one during PID assisted learning. This way, the agent will be exposed to larger parts of the state and action space, increasing exploration.
        
        With PID assistance, the agent should be able to learn a control law that is based on all available information in the system. The reward functions defined in \autoref{eq:CT_reward} and \autoref{eq:VT_reward} are shaped as Gaussian and quadratic functions, and are used for the cross- and vertical-track controllers, respectively. They are shaped differently simply to observe the difference in behavioural outcome, such as the training process and the final tracking error. As explained in \cite{RewardShape}, the Gaussian function is a good candidate for continuous reward functions because of its well defined \textit{reward gradient}. The quadratic function is interesting as it is both negative definite and continuously differentiable. Another attribute of the quadratic function is its relative decrease in punishment as each error approaches zero. Hence, large errors should have a greater influence when performing gradient ascent, and be prioritized during learning.
        
        \begin{equation}
            r_t(\tilde{\chi}, \text{e}, \delta_r) = \alpha_1(1-e^{-5\tilde{\chi}^2}) + \alpha_2(1-e^{-5\text{e}^2})
            + \alpha_3(1-e^{-5\delta_r^2})
            \label{eq:CT_reward}
        \end{equation}
        
        \begin{equation}
            r_t(\tilde{\upsilon}, \text{h}, \delta_s) = \alpha_1 \tilde{\upsilon}^2 
            + \alpha_2 \text{h}^2 
            + \alpha_3 \delta_s^2
            \label{eq:VT_reward}
        \end{equation}
        
        Table \ref{tab:TT_obs} details the elements of the state vector that the agent receives for cross-track control. Note that when learning vertical-track control, elevation and vertical-track error are penalized instead of course and cross-track error. Also, the position of the rudder fin replaces the observation of the elevator fin. The penalization factor for fin actuation is $\alpha_3=\num{-1e-2}$ in both cases.
        
        \begingroup
        \setlength{\tabcolsep}{6pt} 
        \renewcommand{\arraystretch}{1.5} 
        \begin{table}[pos=H]
        \centering
        \caption{Observations made by the agent during PID assisted learning for cross-track control. During vertical-track control, the observation and penalization of the elevation and vertical-track error are interchanged for the course and cross-track error. Furthermore, rudder fin position $\delta_r$ is observed in place of the elevator fin.}
        \begin{tabular}{l c c c}
        \hline
        \textbf{Observation} & & \textbf{Max} & \textbf{$\alpha$} \\
        \hline
        Relative surge speed & $u_{ro} = \frac{u_r}{u_{max}} \in [-1,1]$ & $2$ & 0\\
        Relative sway speed & $v_{ro} = \frac{v_r}{v_{max}} \in [-1,1]$ & $0.3$ & 0\\
        Relative heave speed & $w_{ro} = \frac{w_r}{w_{max}} \in [-1,1]$ & $0.3$ & 0\\
        Roll & $\phi_o = \frac{\phi}{\phi_{max}} \in [-1,1]$ & $\pi$ & 0 \\
        Pitch & $\theta_o = \frac{\theta}{\theta_{max}} \in [-1,1]$ & $\pi$ & 0 \\
        Yaw & $\psi_o = \frac{\psi}{\psi_{max}} \in [-1,1]$ & $\pi$ & 0\\
        Roll rate & $p_o =  \frac{p}{p_{max}} \in [-1,1]$ & 1.2 & 0\\
        Pitch rate & $q_o = \frac{q}{q_{max}} \in [-1,1]$ & 0.4 & 0\\
        Yaw rate & $r_o = \frac{r}{r_{max}} \in [-1,1]$ & 0.4 & 0\\
        Surge error & $\tilde{u}_o = \frac{u_d-u}{u_{max}} \in [-1,1]$ & 2 & 0 \\
        Course error & $\tilde{\chi}_o = \frac{\chi_d - \chi}{\chi_{max}} \in [-1,1]$ & $\pi$ & -2e-2\\
        Elevation error & $\tilde{\upsilon}_o = \frac{\upsilon_d - \upsilon}{\upsilon_{max}} \in [-1,1]$ & $\pi$ & 0 \\
        Cross track error & $e_o = \frac{e}{e_{max}} \in [-1,1]$ & 25 & -5e-2\\
        Vertical track error & $h_o = \frac{h}{h_{max}} \in [-1,1]$ & 25 & 0 \\
        Propeller shaft speed & $n$ & 1 & 0 \\
        Elevator fin position & $\delta_s$ & 1 & 0
        \end{tabular}
        \label{tab:TT_obs}
        \end{table}
        \endgroup

\section{Results and discussion}
\label{section:results}
    In this section we present the major findings of this work. First we present the results for the end-to-end learning, and then for the PID assisted learning.
    
    \subsection{End-to-end learning}\label{section:e2eresults}
     The hyperparameters used during end-to-end learning are given in \autoref{tab:hyperparameters}. The number of steps (T in \autoref{alg:PPO}) and large batch size (M in \autoref{alg:PPO}) lead to long learning times on a desktop computer. 
     
    \begin{table}[pos=h]
    \centering
    \caption{Values of hyperparameters used for training the DRL controllers.}
    \begin{tabular}{c c}\\
    \textbf{Hyperparameter} & \textbf{Value} \\\toprule
    Learning rate & 5e-5 \\  \midrule
    Discount rate & 0.999 \\  \midrule
    GAE parameter & 0.95 \\ \midrule
    \# Actors & 10 \\ \midrule
    \# Steps & 6144\\ \midrule
    Epochs & 4 \\ \midrule
    Batch size & 1024 \\  \midrule
    Min. reward & -500 \\  \midrule
    \end{tabular}
    \label{tab:hyperparameters}
    \end{table}
     
\begin{figure}[pos=H]
        \centering
        \includegraphics[width=1\linewidth]{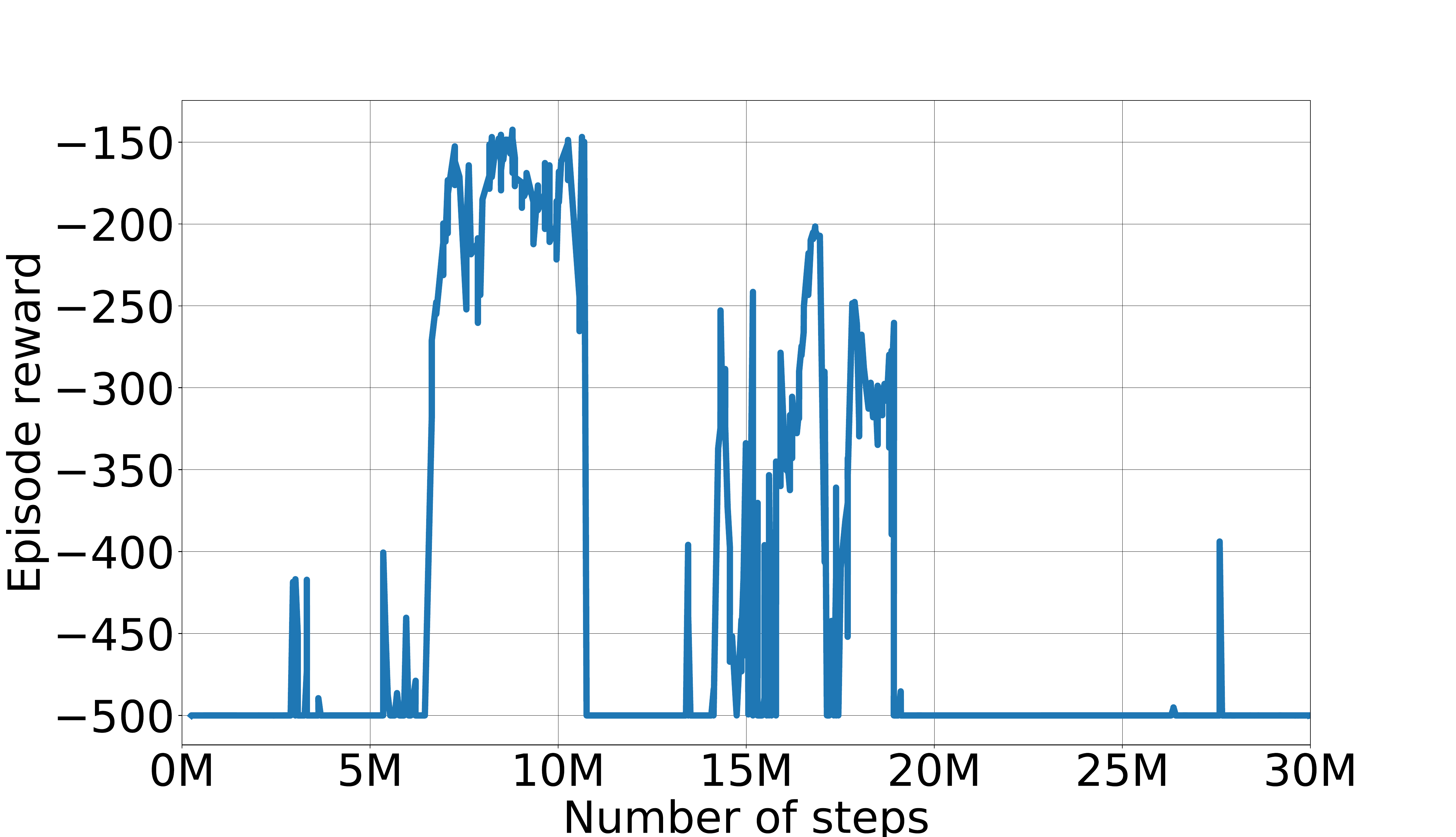}
        \caption{Episode reward when training the end-to-end controller. Performance peaks around 8M time steps.}
        \label{fig:PF_reward}
    \end{figure}

    \autoref{fig:PF_reward} shows the reward value as a function of simulated time steps the AUV has spent exploring the environment. The AUV was left to explore the environment for 30 million time steps, and as seen in \autoref{fig:PF_reward}, the reward peaks at around 8 million. After 10 million time steps an unlearning process is observed. It is not obvious why the agent seemingly unlearns behaviour after 10 million time steps. It is speculated that the agent discovers different possibilities for minimizing the reward function when it receives new information about the environment. The agent might observe a set of unlikely states after 10 million time steps that motivates this new approach. Note that some noise in the learning process is expected, caused by the path being randomly generated for each episode. However, the test results are simulated on a pre-determined path. 
     
\begin{figure}[pos=H]
        \centering
        \includegraphics[width=1\linewidth]{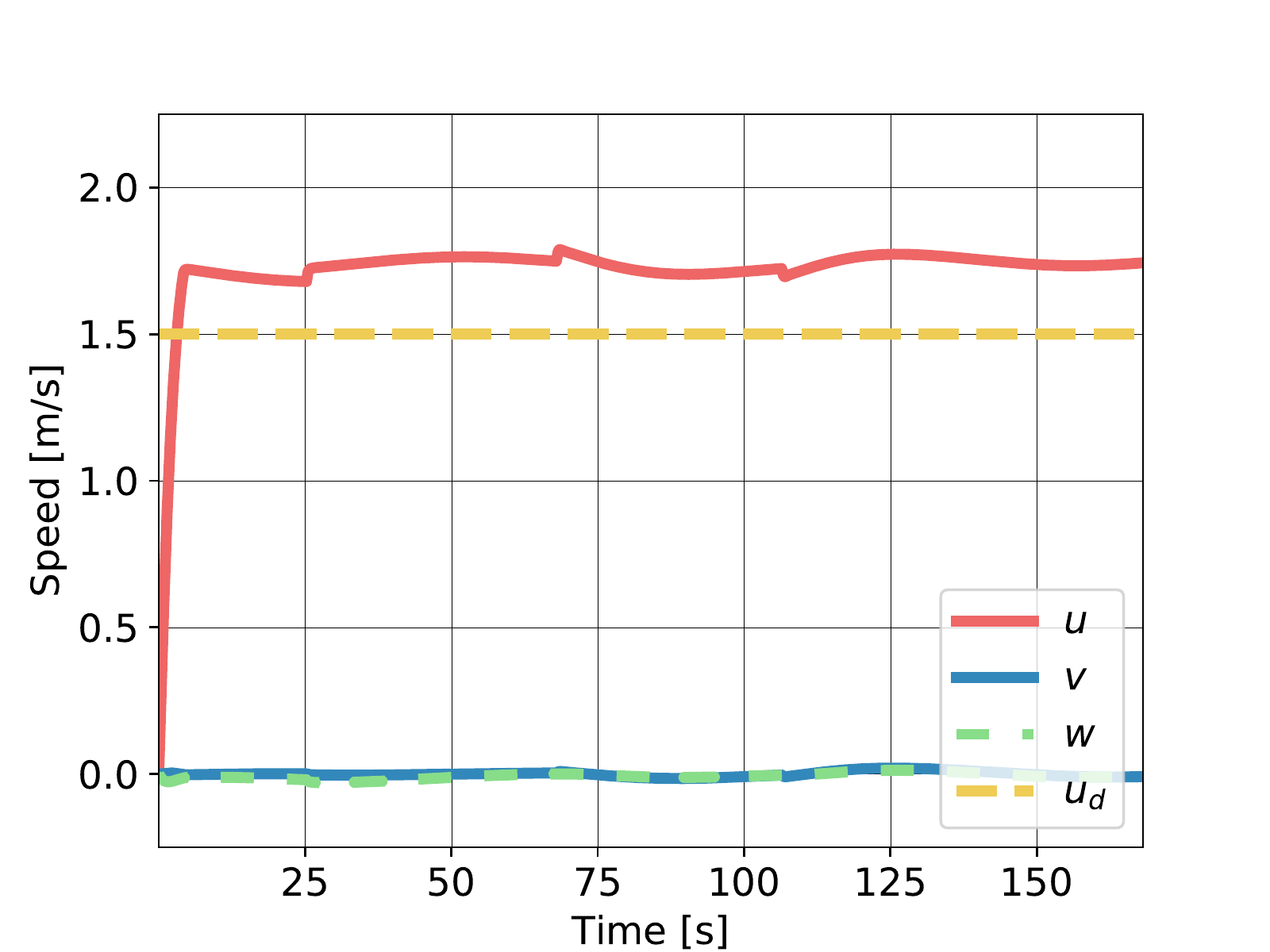}
        \caption{Velocity plot from end-to-end control simulation.}
        \label{fig:PF_vel}
    \end{figure}

\begin{figure}[pos=H]
        \centering
        \includegraphics[width=1\linewidth]{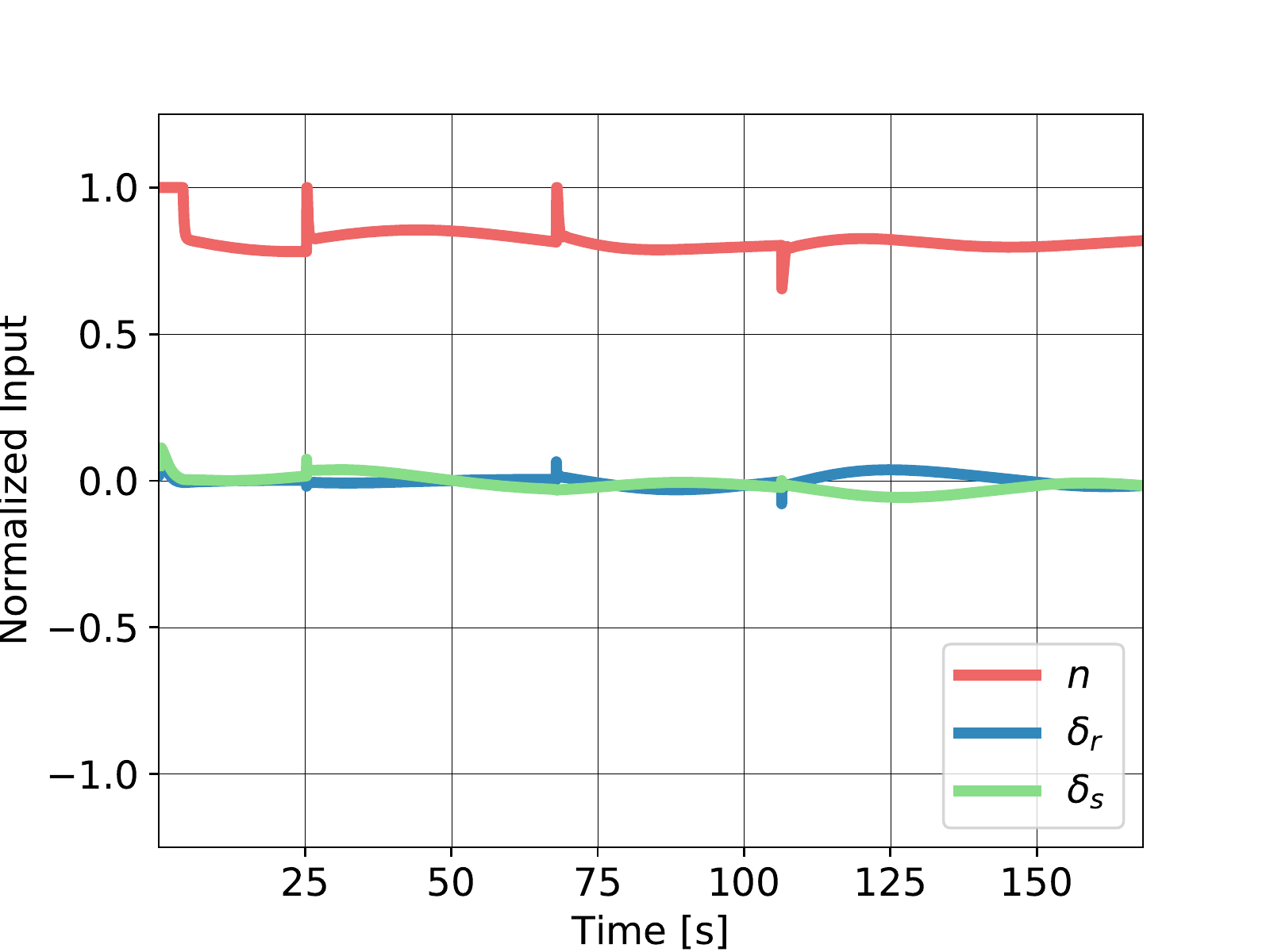}
        \caption{Normalized control inputs for end-to-end control simulation. ($\eta=$~propeller shaft speed, $\delta_r=$~rudder, $\delta_s=$~elevator)}
        \label{fig:PF_control}
    \end{figure}

\begin{figure}[pos=H]
        \centering
        \includegraphics[width=1\linewidth]{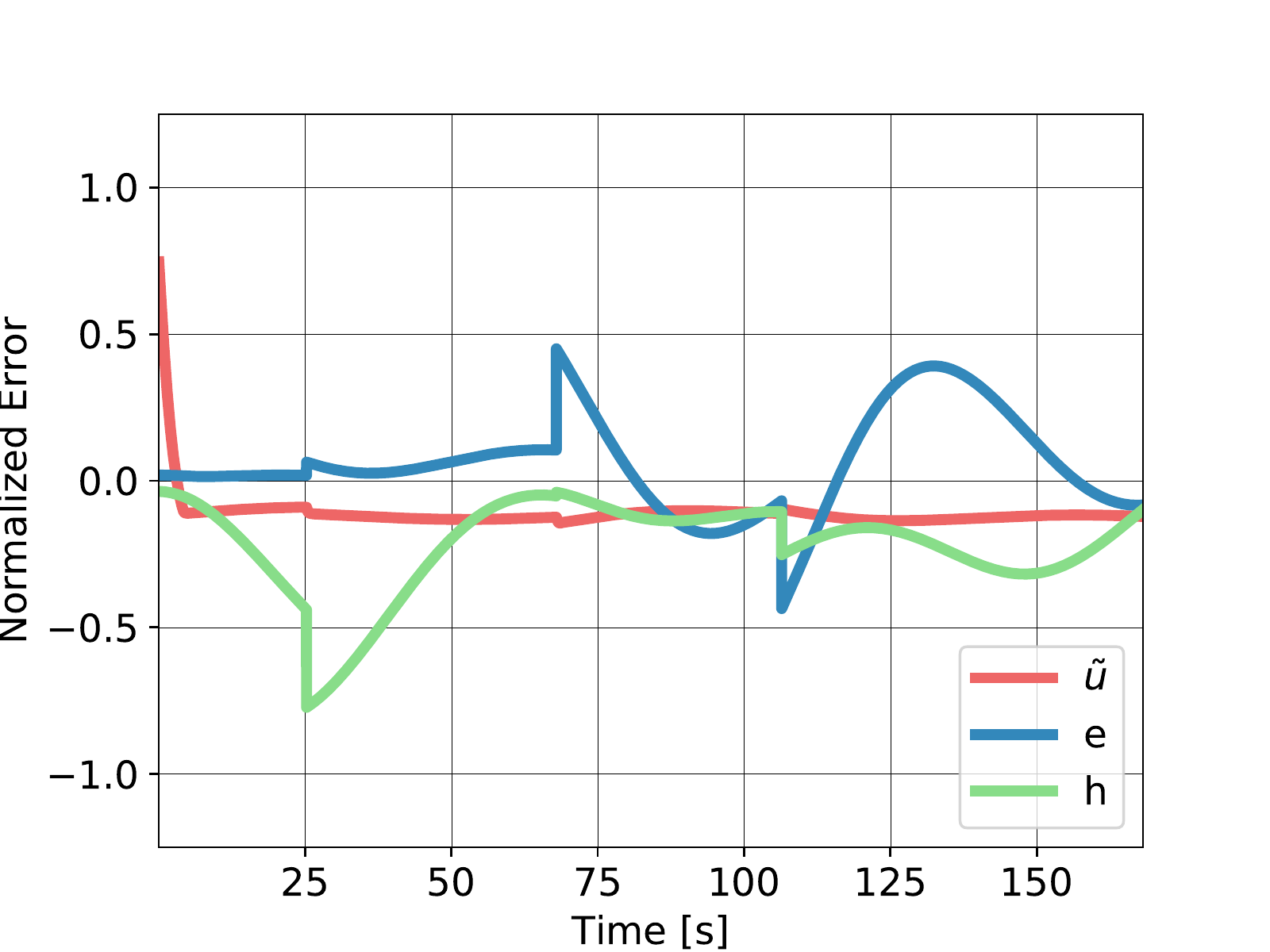}
        \caption{Normalized control errors from test simulations with the end-to-end controller.}
        \label{fig:PF_error}
    \end{figure}

\begin{figure}[pos=H]
        \centering
        \includegraphics[width=1\linewidth, trim={1cm 0cm 2cm 0cm}]{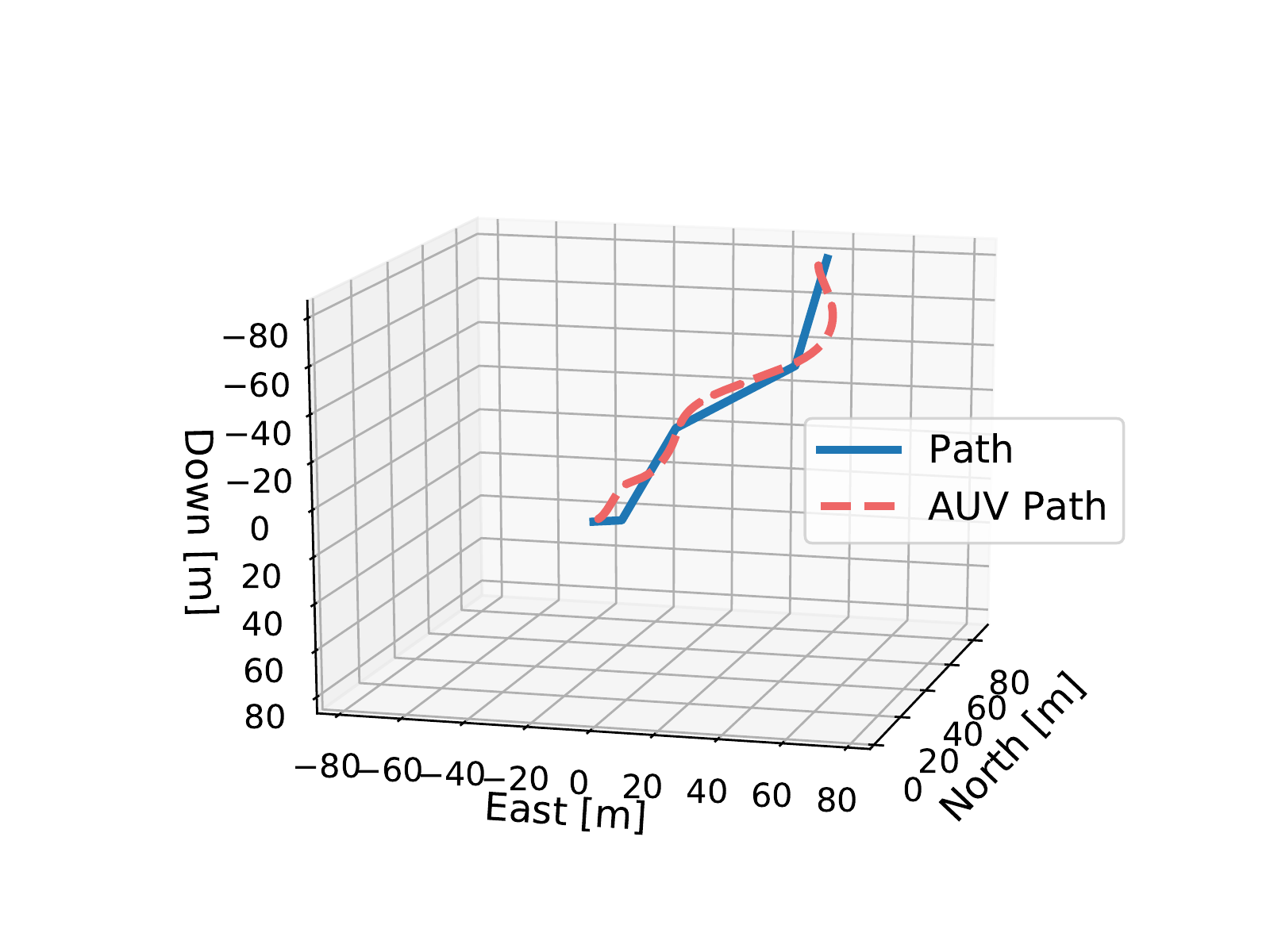}
        \caption{3D plot of the path along with the AUV trajectory for the end-to-end control simulation.}
        \label{fig:PF_3D}
    \end{figure}

     The controller that achieved the highest reward was restored in order to investigate its behaviour after training. Simulations of the AUV with this controller yields the behaviour exemplified in Figure \ref{fig:PF_vel}-\ref{fig:PF_3D}.
     
    \autoref{fig:PF_vel} shows the surge, desired surge, sway and heave speeds when applying the best controller for the end-to-end learning. The RL agent keeps the surge speed close to the setpoint, though an offset is observed. The four abrupt changes in the surge speed happen when the guidance system switches between waypoints.
    
    Next, the normalized control action is seen in \autoref{fig:PF_control}. The propeller thrust is initially high, in order to accelerate the AUV to the desired surge speed. The fin movements are limited, demonstrating the effect of penalizing the angular rates, as discussed in \autoref{subsec:endtoend}. When the guidance system switches between waypoints, the use of control inputs increases.
    
    As with the surge speed, the control errors, seen in \autoref{fig:PF_error}, experience rapid changes when the guidance system switches waypoints. None of the errors are completely eliminated, and an interpretation of the results is that the effect of one actuator disturbs the others, which yields oscillating control errors. 
    
    The effect of changing waypoints is again displayed in \autoref{fig:PF_3D}, which shows a 3D plot of the simulation. When changing waypoints, the AUV overshoots and is unable to fully reduce the errors before the guidance system targets a new waypoint. The performance is not considered as adequate to state that the end-to-end controller solves the path following problem. However, the results are promising and with more research on reward functions and penalization schemes, end-to-end control might be feasible.
    
    The results were obtained by simulating without ocean current disturbances. Better results under ideal conditions are needed before progressing further with this method. 
    
    \begin{figure}[pos=H]
        \centering
        \includegraphics[width=1\linewidth]{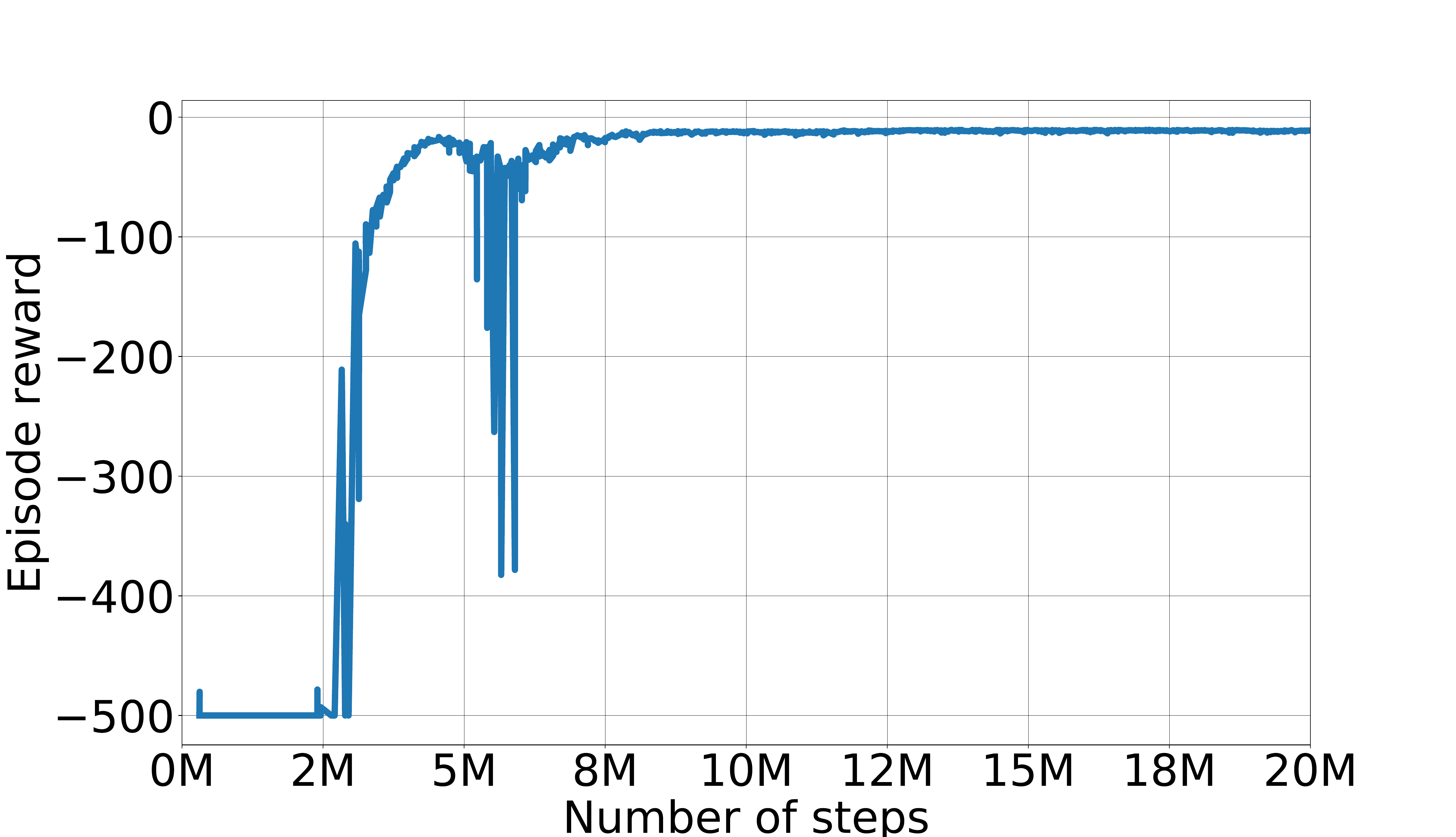}
        \caption{Episode reward when learning cross-track control, where the reward function is given by \autoref{eq:CT_reward}.}
        \label{fig:CT_reward}
    \end{figure}
    
    \subsection{PID assisted learning}\label{section:pid_results}
\begin{figure}[pos=H]
            \includegraphics[width=1\linewidth]{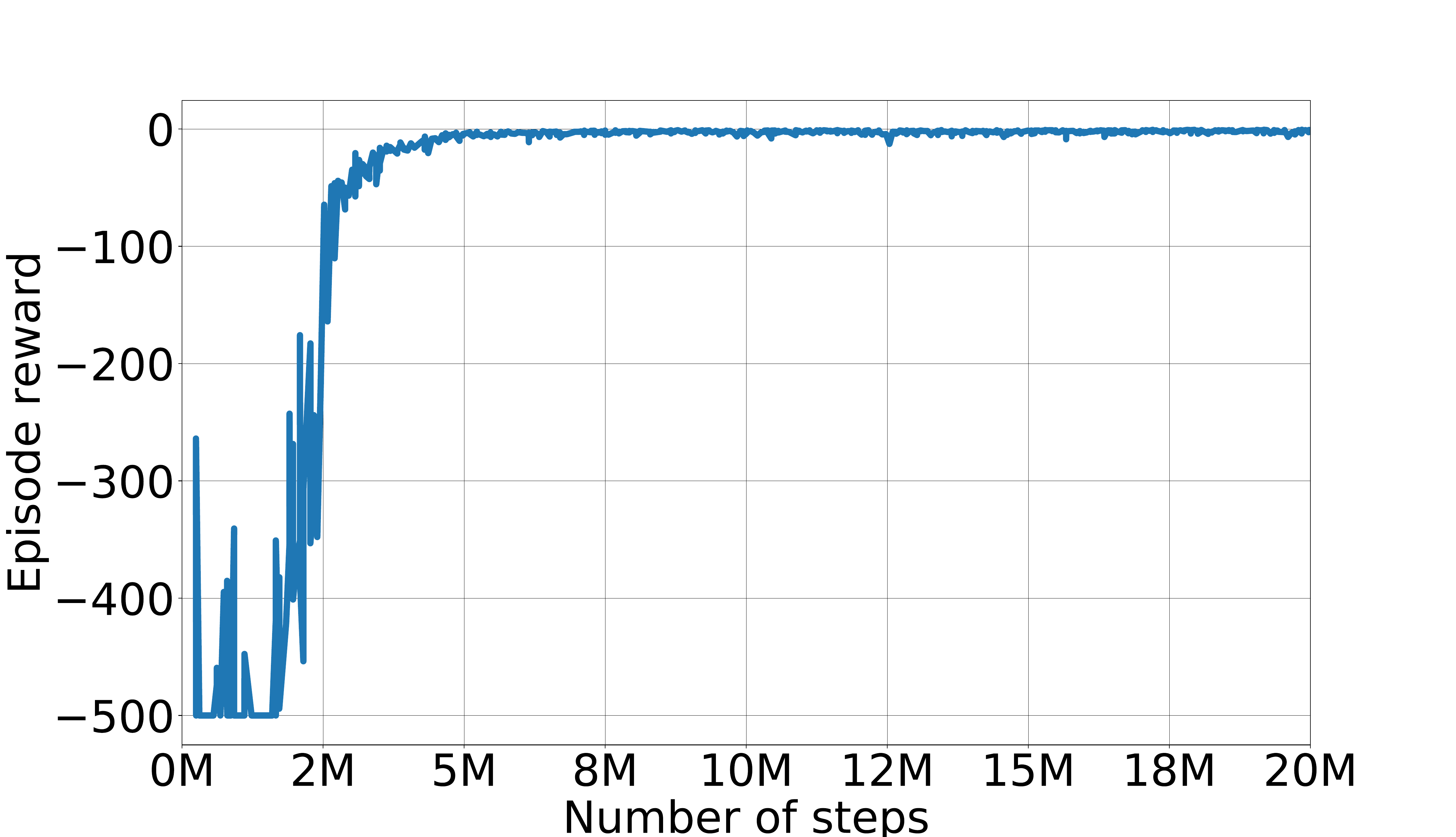}
            \caption{Episode reward when learning vertical-track control, where the reward function is given by \autoref{eq:VT_reward}.}
            \label{fig:VT_reward}
        \end{figure}

         Hyperparameters remained as in \autoref{tab:hyperparameters}, as the results for the end-to-end and PID assisted learning should be comparable. The achieved reward during PID assisted learning is shown in \autoref{fig:CT_reward} and \autoref{fig:VT_reward} when using the reward function from \autoref{eq:CT_reward} and \autoref{eq:VT_reward}, respectively. 
         The episode rewards during training of the cross-track controller is displayed in \autoref{fig:CT_reward}. Compared to \autoref{fig:PF_reward}, the agents learning seems faster and more stable, as intended. In this learning scenario, the reward function is simpler and as the agent is only learning one control input at a time, the state space being explored is consequently smaller.  There is a small dip in accumulated reward at 6M time-steps, but this is quickly solved and the learning saturates at 10M time-steps. Hence, there are no signs of the unlearning behaviour that was observed in the end-to-end case.
        
         In \autoref{fig:VT_reward}, the episodic reward during training of the vertical-track controller is seen. Here, the agent experiences no drop in reward and the performance saturates at about 5M time-steps, indicating faster learning. This leads us to believe that the quadratic reward function from \autoref{eq:VT_reward} might be better suited for this control problem. 
        
        The PID assisted learning scheme with the quadratic reward function reduces the learning time by at least $66\%$ compared to the end-to-end trained controller. This can be stated with confidence on the basis that after 30M timesteps the latter controller's performance is not adequate, as seen in \autoref{fig:PF_3D}. PID assisted learning yields a very capable controller after 10M time-steps. This is also highlighted in the simulation results, which indicate significantly better performance on path following after combining the cross and vertical-track MLC obtained with PID assisted learning. 
        
        Note that the PID controllers are only necessary when training, but the PI controller for surge speed is used in the simulation results in \autoref{section:pid_results}. 
        There is no reason to doubt whether a velocity controller could learn equivalently to the tracking controllers, i.e. by PID assistance. However, this is not pursued as earlier experiments (not included in this article) showed that perfect velocity tracking is achievable through end-to-end learning for velocity control. 
        
        To best be able to compare the reward functions and observe their effect on learning, the disturbing current was removed during training. Additionally, the path generated in all test simulations are identical. Regardless of the networks training without disturbances, the simulation results show great performance on the path-following problem, both with and without an ocean current perturbing the system. 
        The performance of the combined PID assisted controllers after training is presented in the next section.
        
        \subsubsection{Simulation Results without Current Disturbance}\label{sec:TTSR1}
\begin{figure}[pos=H]
                \includegraphics[width=1\linewidth]{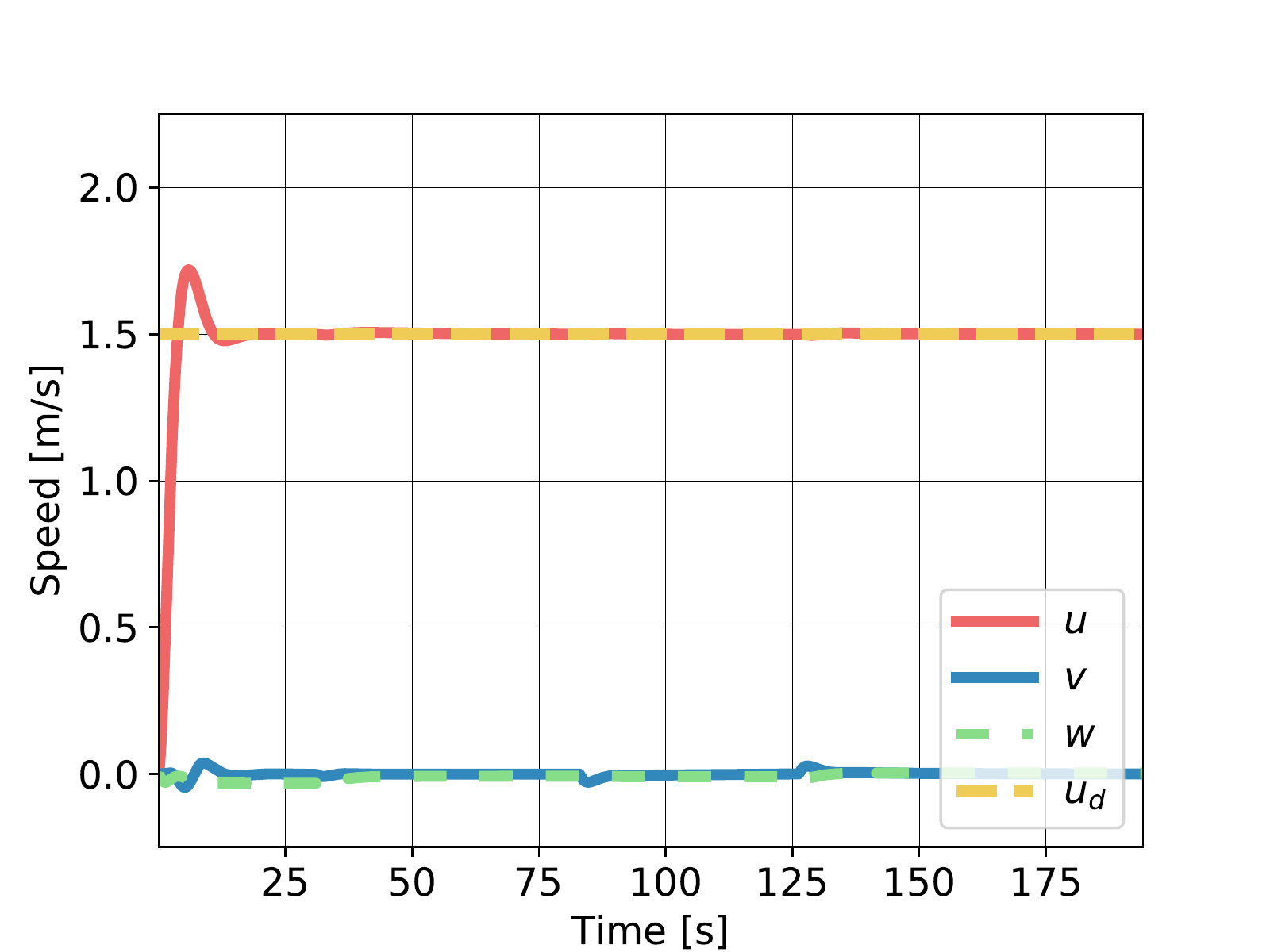}
                \caption{Velocity plot from PID assisted control simulation.}
                \label{fig:TT1_vel}
            \end{figure}

\begin{figure}[pos=H]
                \includegraphics[width=1\linewidth]{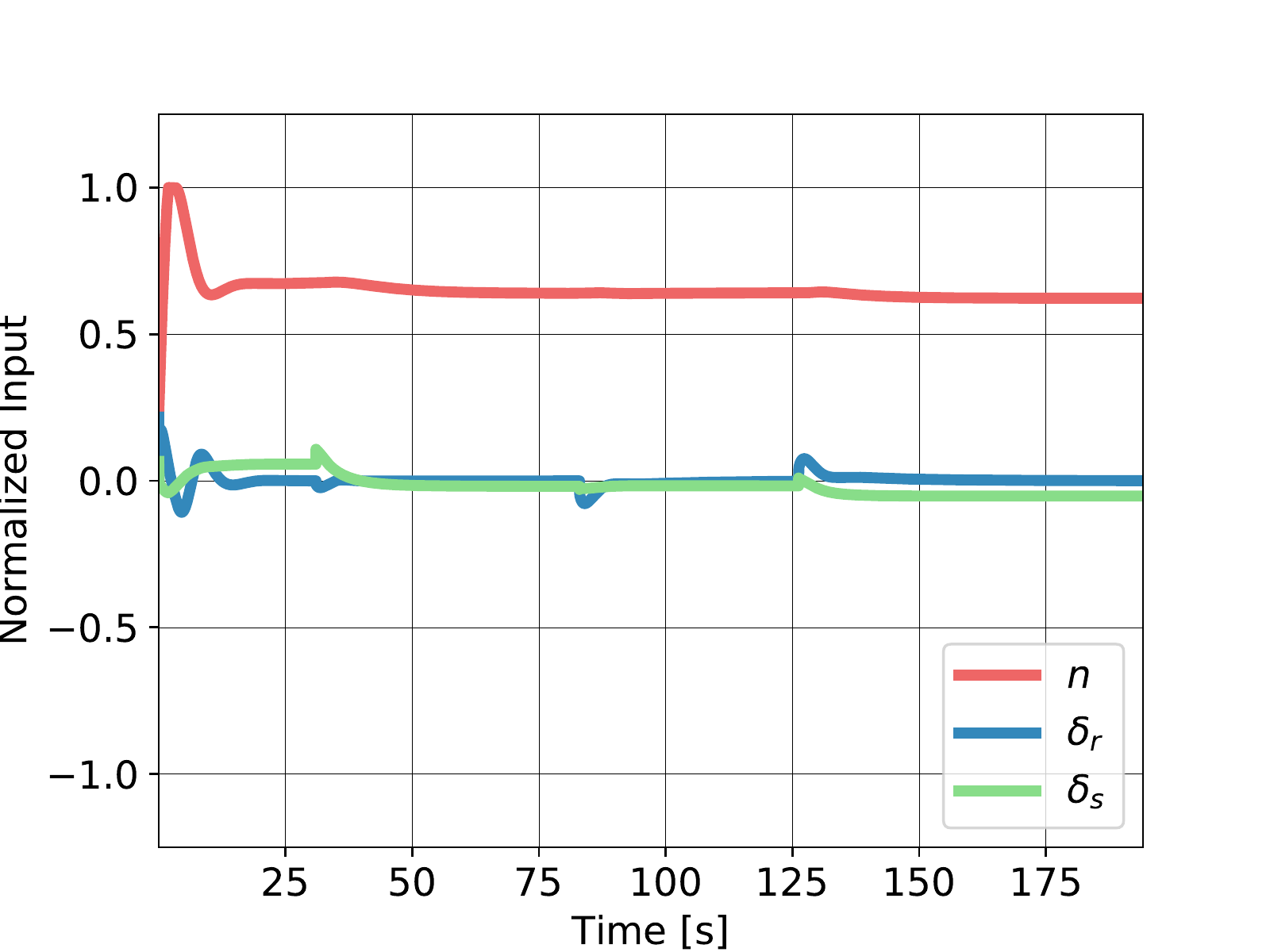}
                \caption{Normalized control inputs from PID assisted control simulation.}
                \label{fig:TT1_control}
            \end{figure}

\begin{figure}[pos=H]
                \includegraphics[width=1\linewidth]{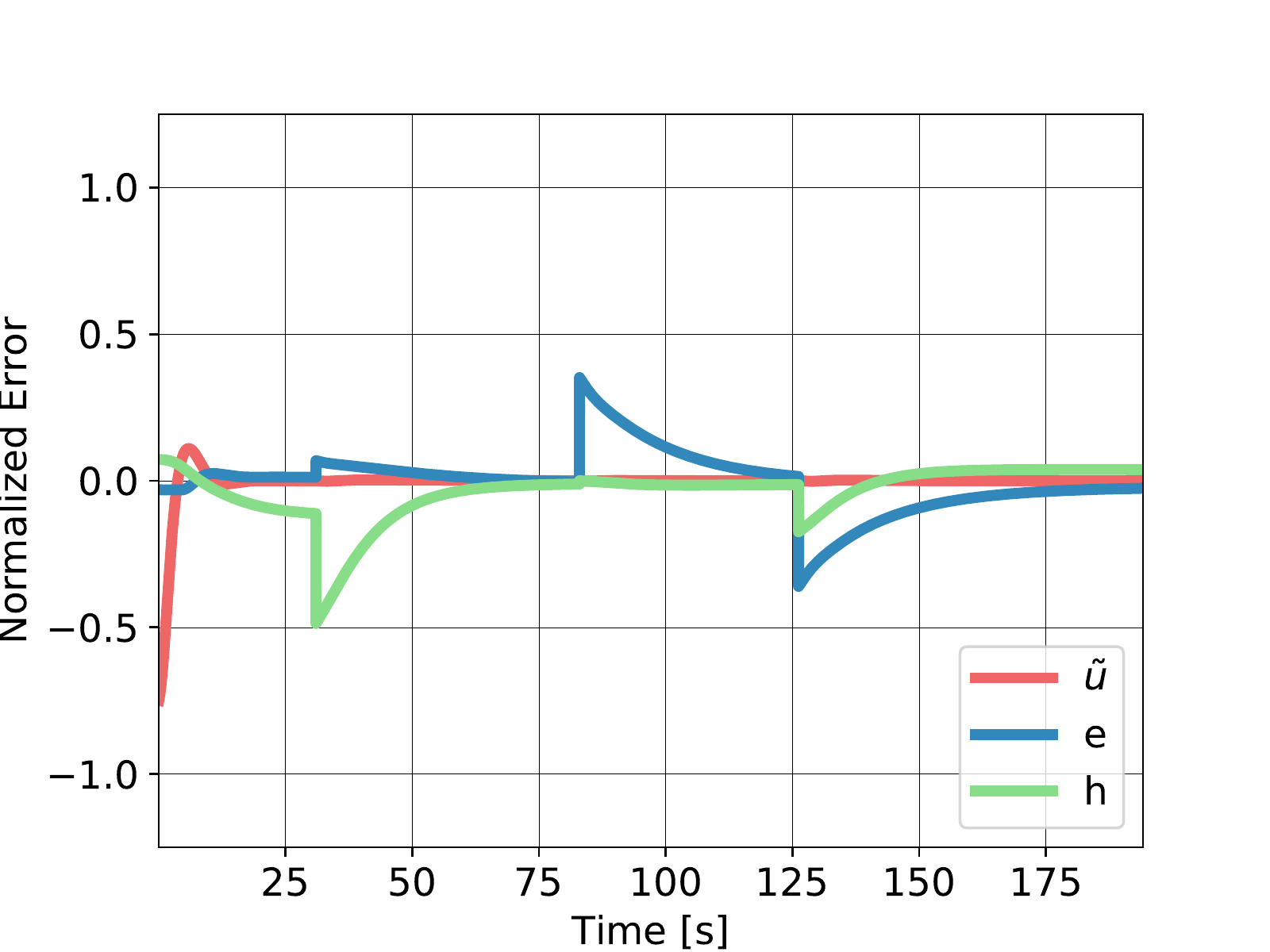}
                \caption{Normalized control errors from the PID assisted control simulation. All control errors are eliminated and the deviations are only the result of switching waypoints.}
                \label{fig:TT1_error}
            \end{figure}

\begin{figure}[pos=H]
                \includegraphics[width=1\linewidth, trim={1cm 0cm 2cm 0cm}]{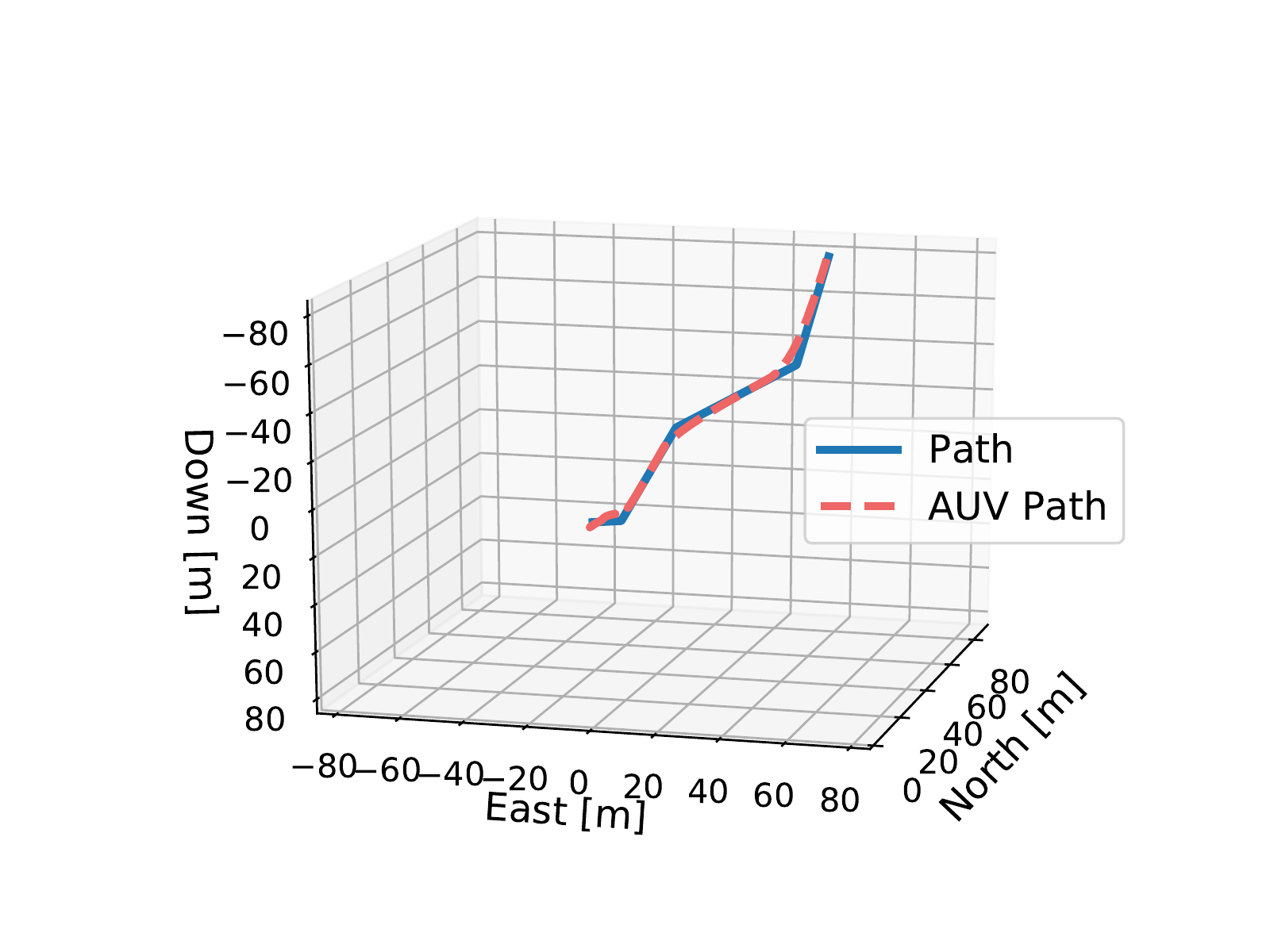}
                \caption{3D Plot of the path along with the AUV trajectory for the PID assisted control simulation.}
                \label{fig:TT1_3D}
            \end{figure}

            In addition to a greatly reduced training time, \autoref{fig:TT1_vel} to \ref{fig:TT1_3D} reveal better tracking performance than the end-to-end trained controller.
            
            In \autoref{fig:TT1_vel} we observe a typical PI setpoint regulation of the surge speed. The effects from switching waypoints on surge speed are negligible, which indicates that the disturbances from the control fins are small. 
            
            The normalized control action is pictured in \autoref{fig:TT1_control}. The fins are used conservatively, as intended by the penalization term in the reward functions. The control is smooth and well behaved and the oscillatory, aggressive behaviour seen in the end-to-end simulation is not displayed here. Furthermore, the effect of waypoint switching is minuscule, in accordance with the negligible effects observed in the velocity plot. 
            
            Control errors stay very close to zero, and only see an increase when the guidance system targets new waypoints, as seen in \autoref{fig:TT1_error}. Perfect surge speed regulation is observed, and the DRL controllers make the AUV follow the path accurately. 
            
            The increase in performance is truly reflected in \autoref{fig:TT1_3D}, where the trajectory of the AUV and the path are displayed. The increase in performance is self-evident when comparing this plot with \autoref{fig:PF_3D}. There is no longer an overshoot, and the AUV stays close to desired path, even when changing waypoints. 
            
        \subsubsection{Simulation Results with Current Disturbance}\label{sec:TTSR2}
            
            This section presents the simulation results when employing the same controllers from \hyperref[sec:TTSR1]{Section 5.2.1} in the presence of ocean currents. The direction of the current is randomly initialized, while the intensity is simulated as a random walk within the interval $0.5$ to $1.0~ms^{-1}$.
            
\begin{figure}[pos=H]
                \includegraphics[width=1\linewidth]{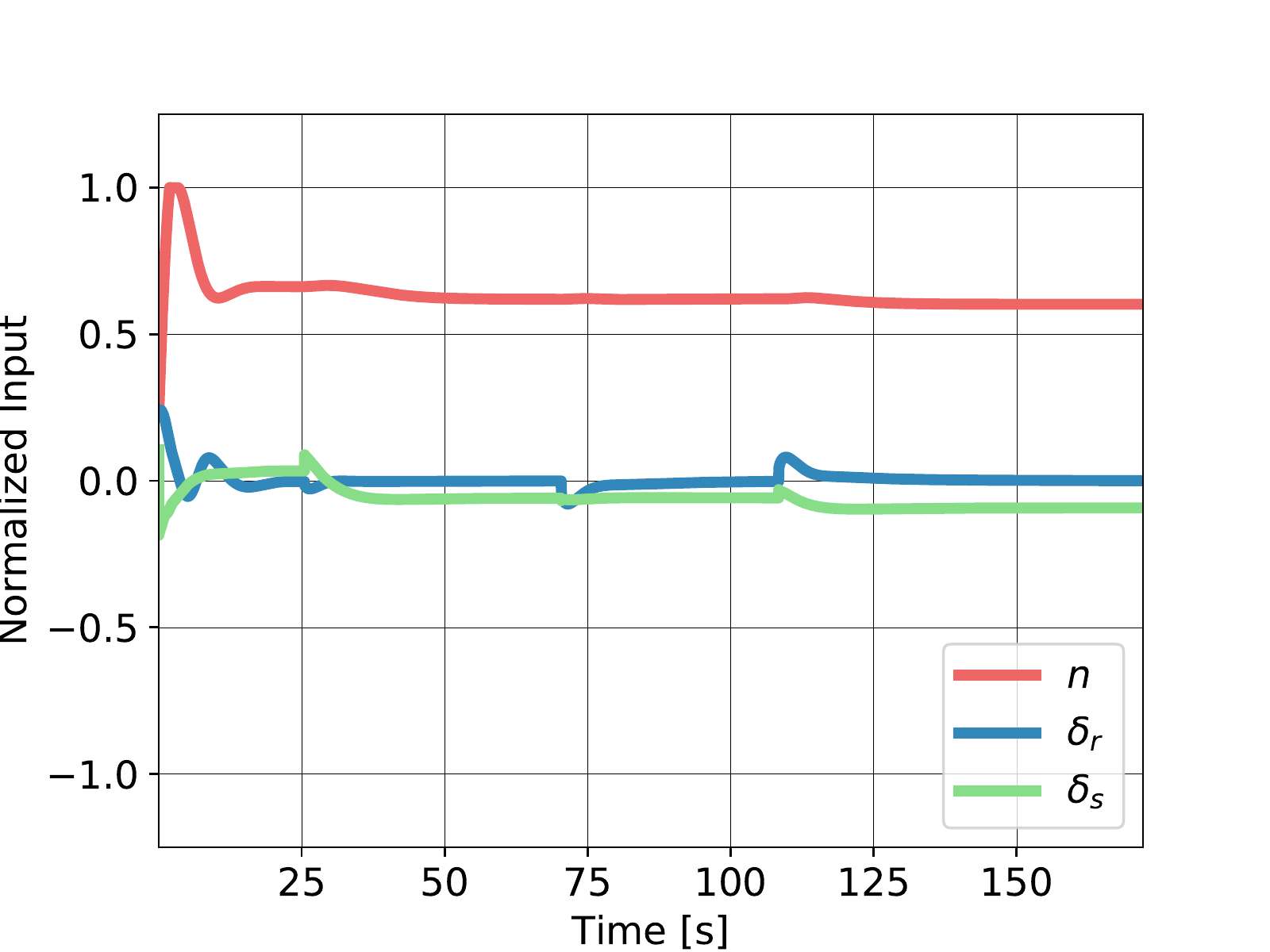}
                \caption{Normalized control input from the PID assisted control in the presence of an ocean current.}
                \label{fig:TT2_control}
            \end{figure}

            When ocean currents are present, the DRL controllers make greater use of the fins, as seen in \autoref{fig:TT2_control}. Especially the elevator fin is utilized more than before, due to the current having a large vertical component. This does not seem to introduce unwanted behaviour.
            
\begin{figure}[pos=H]
                \includegraphics[width=1\linewidth]{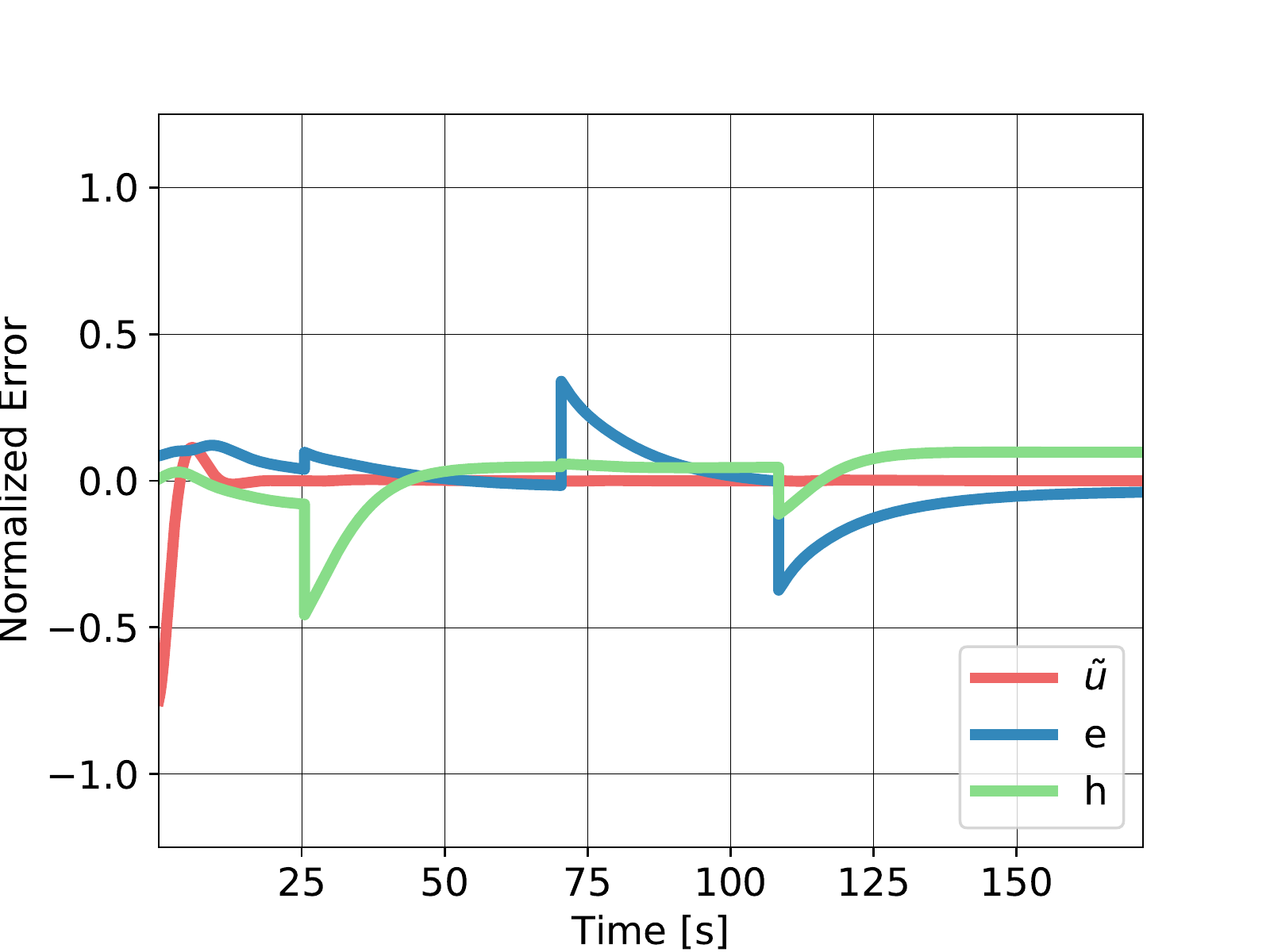}
                \caption{Normalized control errors from the PID assisted control simulation in the presence of an ocean current.}
                \label{fig:TT2_error}
            \end{figure}

            \autoref{fig:TT2_error} presents the control errors, which reveals a slight offset in vertical tracking. The agent has inferred that it must increase the control input in order to compensate for the current, but since it has not experienced this scenario during training, it is not able to compensate completely. However, it is an encouraging finding that the AUV is able to perform so well in the presence of  an ocean current with varying intensity when it has been trained under perfect conditions.
            
\begin{figure}[pos=H]
                \includegraphics[width=1\linewidth, trim={1cm 0cm 2cm 0cm}]{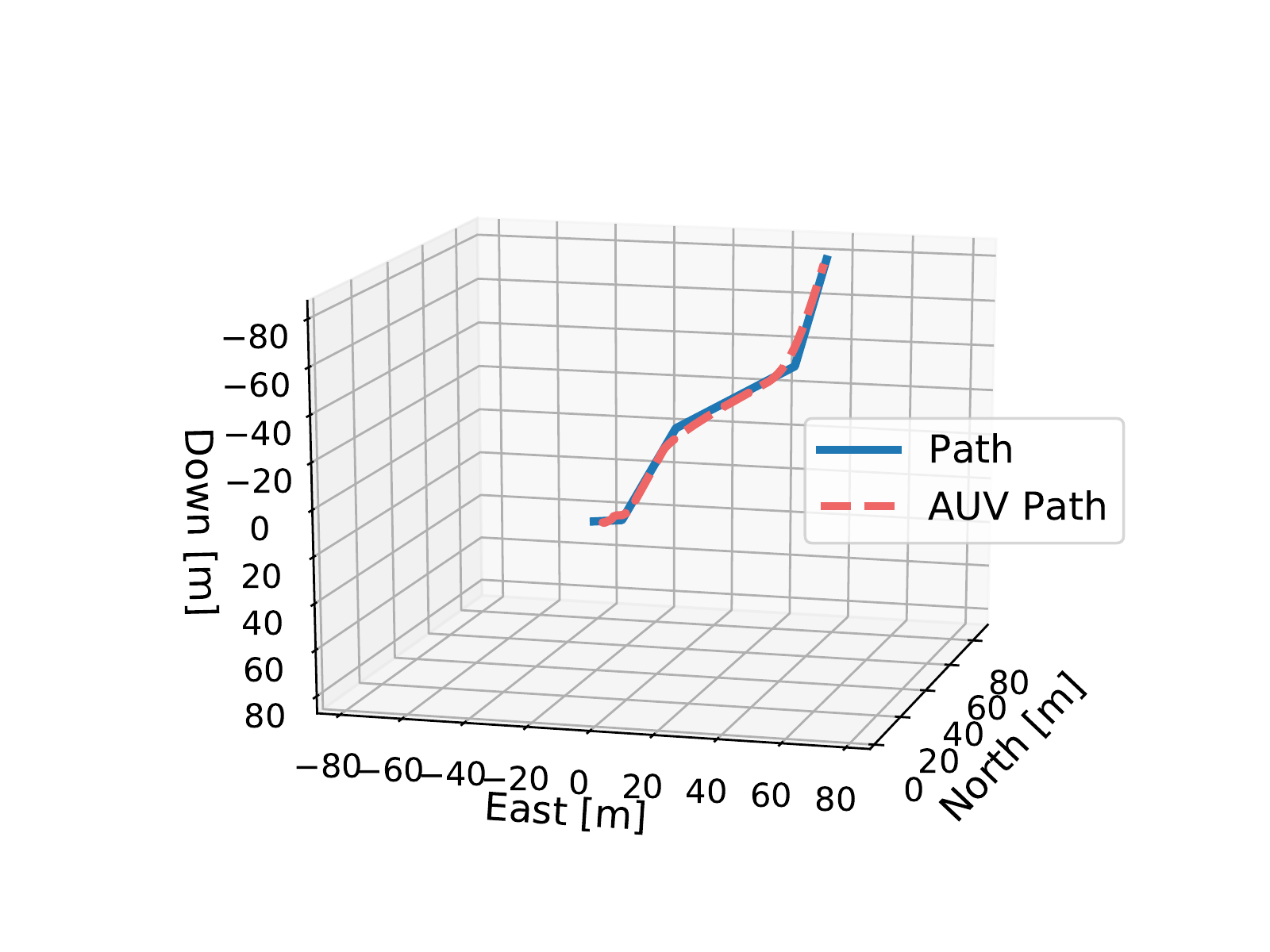}
                \caption{The 3D plot shows that the tracking is still performing well, but the stationary deviation in the vertical component seen in \autoref{fig:TT2_error} is noticed here as well.}
                \label{fig:TT2_3D}
            \end{figure}

            In \autoref{fig:TT2_3D}, a 3D plot of the simulation is seen. The constant control deviation is noticeable, but as implied by the input and error plots the behaviour is still impressive. 
            
            As previously stated, the neural networks were trained without any currents or disturbances, so it is not expected to achieve perfect tracking in these simulations. The main idea is to demonstrate that the controllers still, to a great extent, achieve path following, even when exposing them to previously unseen scenarios. Overall these results can be seen as a form of robustness testing, and the results are quite exciting.
            
            By adding disturbances during training, robustness could be increased further. In that way, it would be possible to learn to compensate for the currents by exploration and exploitation. 

\section{Conclusions and future work}
\label{section:conclusion} In this section we present the main conclusions including the strengths and weaknesses of the current work, as well as suggestions for further research.

    The 3D path following problem for AUVs was investigated by utilizing DRL controllers, more specifically with the PPO algorithm. Two different approaches were considered: end-to-end learning, where the agent is left entirely alone to explore the solution space in its search for an optimal policy, and PID assisted learning, where the DRL controller is essentially split into three separate parts, each controlling its own actuator (rudder, elevator and thrust). When training one actuator at a time, the two others are being controlled by stable PID controllers such that the control objectives can be achieved. End-to-end learning showed promising results, but due to a highly complex reward scheme and sensitivity to training set-up, more research is needed to obtain better performance with this method. The study of DRL in practical applications are mostly based on heuristics and experimental data, which prompts major challenges in the fusion of cybernetics and AI. PID assisted learning gave excellent simulation results, and together with the quadratic cost function it was found to significantly reduce the number of simulated time-steps needed to train the controller. Furthermore, an advantage with the DRL approach, compared to traditional control methods, is the lack of need for an accurate underlying world model to achieve excellent results. No tuning is needed, it does that itself by exploration and exploitation. No a priori information is needed, other than that the traditional control methods (PID) are able to stabilize subprocesses during learning. The results make us hopeful that extending the aforementioned methods to incorporate collision avoidance, can take AUV motion control systems one more step towards true autonomy. The performance of the trained agent for unseen scenarios therefore demonstrate the robustness of DRL for discovering generalizing control strategies.
 
    We foresee several potential research directions to expand on the current work. Firstly, we got little insight into the learning process due to the involvement of a black-box DNN. In order for MLC to be projected as an alternative to traditional controllers, the issue of interpretability and explainablity will have to be addressed. Some inspiration can be taken from our recent work \cite{vaddireddy2020fes} where we employed symbolic regression to discover hidden physics from data. It remains to be seen how similar approaches can be used to discover efficient control laws. Another natural extension of the work will be to train the model for attaining the dual objective of collision avoidance and path following. Some encouraging results in this regard have been reported in \cite{meyer2020colreg} and \cite{meyer2020taa}, albeit for surface vehicles. Extension of the work to 3D with 6DOF and explanation of the training process will go a long way in making DRL palatable for safety critical autonomous systems.   

    Towards the end, we stress that our paper lays the groundwork for further research, which may, given equally positive results, bring significant value to the field of autonomous guidance. 

\section*{Acknowledgment}
The authors acknowledge the financial support from the Norwegian Research Council and the industrial partners: DNV GL, Kongsberg and Maritime Robotics of the Autosit project. (Grant No.: 295033).
\bibliographystyle{cas-model2-names}
\bibliography{cas-refs}
\bio{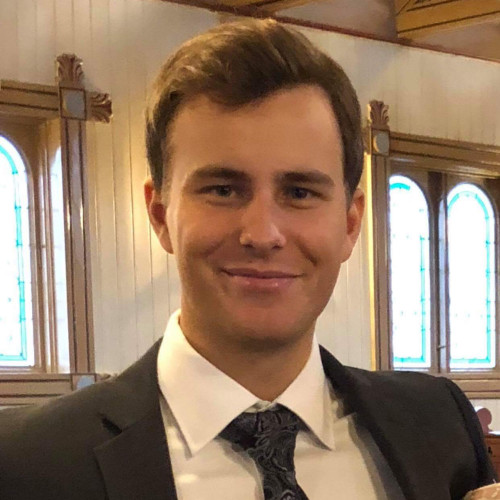}
Simen Theie Havenstrøm is currently pursuing a Masters degree in Cybernetics and Robotics at the Norwegian University of Science and Technology. He received a Bachelors degree in Automation and Industrial IT from the University of South-Eastern Norway, and is a certified automation technician with offshore experience from the Norwegian oil industry. 
\endbio
\vskip3pt
\bio{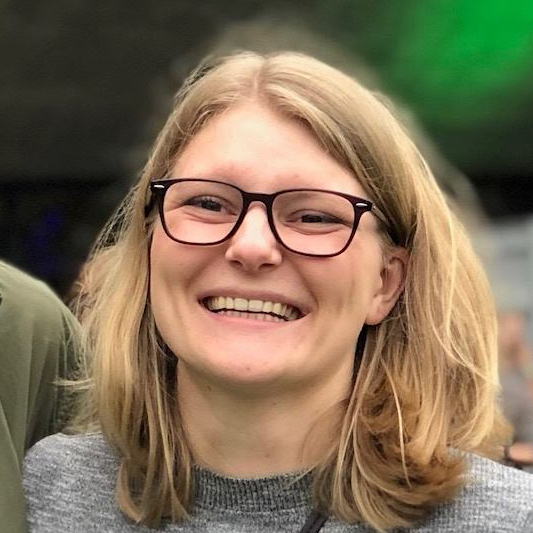}
{Camilla Sterud} holds a research scientist position in the Department of Mathematics and Cybernetics at SINTEF Digital. Prior to joining SINTEF she received a Masters degree in Cybernetics and Robotics in 2019 from NTNU. Her broad interest is in the area of artificial intelligence and machine learning applied to engineering applications.   
\endbio
\vskip3pt
\bio{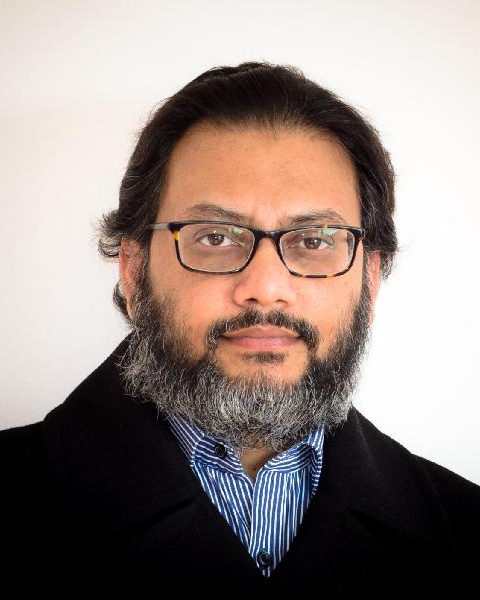}
Adil Rasheed is the professor of Big Data Cybernetics in the Department of Engineering Cybernetics at the Norwegian University of Science and Technology where he is working to develop novel hybrid methods at the intersection of big data, physics driven modeling and data driven modeling in the context of real time automation and control. He also holds a part time senior scientist position in the Department of Mathematics and Cybernetics at SINTEF Digital where he led the Computational Sciences and Engineering group between 2012-2018. He holds a PhD in Multiscale modeling of Urban Climate from the Swiss Federal Institute of Technology Lausanne. Prior to that he received his bachelors in Mechanical Engineering and a masters in Thermal and Fluids Engineering from the Indian Institute of Technology Bombay.
\endbio
\vskip3pt
\bio{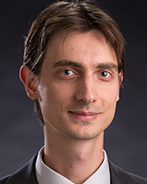}
Omer San received his bachelors in aeronautical engineering from Istanbul Technical University in 2005, his masters in aerospace engineering from Old Dominion University in 2007, and his Ph.D. in engineering mechanics from Virginia Tech in 2012. He worked as a postdoc at Virginia Tech from 2012-'14, and then from 2014-'15 at the University of Notre Dame, Indiana.  
He has been an assistant professor of mechanical and aerospace engineering at Oklahoma State University, Stillwater, OK, USA, since 2015. He is a recipient of U.S. Department of Energy 2018 Early Career Research Program Award in Applied Mathematics. His field of study is centered upon the development, analysis and application of advanced computational methods in science and engineering with a particular emphasis on fluid dynamics across a variety of spatial and temporal scales. 
\endbio

\end{document}